\documentclass[aps,twocolumn,pra,superscriptaddress,amsmath,showpacs,tightenlines,pdflatex,longbibliography]{revtex4-2}
\usepackage{amssymb}
\usepackage{amsmath}
\usepackage{dcolumn}
\usepackage{graphicx}
\usepackage{subfigure}
\usepackage{mathrsfs}
\usepackage{appendix}
\usepackage{graphicx}
\usepackage{booktabs}
\usepackage{color}
\usepackage{bm}
\usepackage{url}
\usepackage[colorlinks]{hyperref}
\hypersetup{%
    plainpages=true,
    breaklinks=true,
    hypertexnames=false,
    pageanchor=true,
    colorlinks=true,
    linkcolor={blue},
    citecolor={blue},
    urlcolor={blue},
    anchorcolor={black}
}

\begin{document}

\title{Quantum gyroscope based on the cavity magnomechanical system}
\author{Zhe-Qi Yang}
\affiliation{Fujian Key Laboratory of Quantum Information and Quantum Optics, College of Physics and Information Engineering, Fuzhou University, Fuzhou, Fujian 350108, China}
\author{Lei Chen}
\affiliation{Fujian Key Laboratory of Quantum Information and Quantum Optics, College of Physics and Information Engineering, Fuzhou University, Fuzhou, Fujian 350108, China}
\author{Yu-Rong Lin}
\affiliation{Fujian Key Laboratory of Quantum Information and Quantum Optics, College of Physics and Information Engineering, Fuzhou University, Fuzhou, Fujian 350108, China}
\author{Wei Qin}
\affiliation{Center for Joint Quantum Studies and Department of Physics, School of Science, Tianjin University, Tianjin 300350, China}
\author{Zhi-Rong Zhong}
\email{zhirz@fzu.edu.cn}
\affiliation{Fujian Key Laboratory of Quantum Information and Quantum Optics, College of Physics and Information Engineering, Fuzhou University, Fuzhou, Fujian 350108, China}

\date{\today }

\begin{abstract}
High-precision rotational angle measurement in noise-prone environments holds critical importance in aerospace engineering, military navigation, and related domains. In this paper, we propose a quantum gyroscope scheme based on a cavity magnomechanical system, which enables high-precision rotation angle detection by harnessing hybrid light-magnon interactions. Central to this framework is the employment of a two-mode squeezed coherent state, generated via parametric coupling of dual quantized optical fields with collective spin excitations (magnons), serving as the quantum metrological probe. We demonstrate that this scheme can significantly reduce quantum noise to levels far below the shot-noise limit. Furthermore, in the non-Markovian case, the performance of the quantum gyroscope in a dissipative environment does not deteriorate over time, provided that the environmental spectral density satisfies certain conditions. These findings provide critical insights for advancing miniaturized quantum gyroscopes with sub-microradian precision, addressing long-standing challenges in inertial navigation systems under strong ambient noise.

\end{abstract}

\maketitle

\section{INTRODUCTION}

Quantum precision measurement, a crucial area of research in contemporary quantum information science, aims to surpass classical measurement limits by utilizing quantum resources. Based on the Sagnac effect \cite{RevModPhys.39.475}, gyroscopes have been experimentally demonstrated in various platforms, including optical interferometric systems \cite{Lai2020,PhysRevLett.125.033605}, ultracold matter-wave systems \cite{PhysRevLett.78.760,PhysRevLett.97.240801,PhysRevLett.114.140404,Campbell_2017}, and acoustic domains \cite{PhysRevApplied.22.014061}. The measurement precision of classical Sagnac gyroscopes is significantly affected by environmental noise, and theoretically, their sensitivity is limited by the shot-noise limit (SNL). To improve the precision and sensitivity of gyroscopes, several strategies have been proposed. Matter-wave gyroscopes exploit the de Broglie wavelength scaling to achieve area-normalized sensitivities surpassing optical counterparts  by serveal orders of magnitude \cite{PhysRevA.57.4736,PhysRevLett.78.2046,savoie2018interleaved}. However, their low bandwidth and limited operational lifetime restrict their broader applicability \cite{PhysRevLett.116.183003}. A hybrid strategy that combines mechanical and atomic gyroscope has been suggested to overcome these limitations, but this approach is still constrained by the SNL \cite{10.1063/1.4897358,PhysRevApplied.10.034030}.

With the rapid development of quantum information technology, quantum gyroscopes have emerged as a promising solution for achieving ultra-high sensitivity. Capitalizing on quantum resources such as squeezed light states \cite{PhysRevD.23.1693,PhysRevLett.118.140401,PhysRevResearch.1.032024} and multiparticle entangled systems \cite{PhysRevLett.112.103604,doi:10.1126/science.aag1106}, quantum gyroscopes can enhance measurement precision, surpass the SNL, and potentially approach the Heisenberg limit. This concept has stimulated significant research activity, resulting in a variety of implementation schemes that illustrate the feasibility of quantum-enhanced gyroscopic operation through various physical platforms \cite{Mehmet:10,Jiao:23,10.1063/5.0135084,10.1063/1.5066028,PhysRevApplied.14.034065,Fink_2019,s20123476,RevModPhys.90.035005,10.1063/5.0050235}. Notable implementations cover several areas. Phase-sensitive injection protocols suppress shot noise using squeezed vacuum states \cite{Mehmet:10}. Robust two-mode squeezing configurations maintain super-Heisenberg scaling even in decoherence-prone environments \cite{Jiao:23}. Furthermore, hybrid interferometric architectures that integrate SU(1,1) state engineering with Sagnac-type configurations for enhanced rotational sensitivity have been explored \cite{10.1063/5.0135084}. Collectively, these approaches demonstrate the versatility of quantum metrological techniques in overcoming conventional detection limits. However, the superiority of quantum gyroscopes, particularly in achieving Heisenberg-limited sensitivity for rotational sensing, remains largely unrealized in practical  implementations \cite{RevModPhys.90.035005,10.1063/5.0050235}. A significant challenge is the inevitable decoherence caused by noise in the microscopic world, which degrades the quantum states and leads to reduced measurement precision and stability. Studies have demonstrated that quantum gyroscopes utilizing squeezed states \cite{PhysRevA.81.033819,PhysRevA.95.053837} and entangled states \cite{PhysRevLett.102.040403,PhysRevA.80.013825,PhysRevLett.107.083601} quickly lose their advantages when environmental noise and photon loss are taken into account. In fact, their measurement precision may return to or even become worse than the SNL under these conditions. Thus, developing a practically viable quantum gyroscope with high stability and precision remains a significant challenge in current research.

Magnons, the quantized spin wave excitations associated with spin ordering in magnets, have received increasing attention due to their capability for carrying, transporting and processing quantum information \cite{YUAN20221,PhysRevA.103.043704,LENK2011107,PhysRevA.103.063708}. Cavity magnomechanical systems represent a novel class of tripartite quantum platforms that integrate high-Q optical resonators, magnetostrictive materials, and nanomechanical oscillators. These systems establish coherent energy exchange channels among magnonic, photonic, and phononic subsystems through optomagnonic and magnomechanical interactions \cite{doi:10.1126/sciadv.1501286,PhysRevX.11.031053,PhysRevApplied.13.064001}. Furthermore, they aim to utilize the interaction between cavity photons and magnons to facilitate research in spintronics \cite{Chumak2015} and quantum technologies \cite{Sarma_2021,PhysRevB.109.L041301}. Thanks to the precise control over the magnon frequency and coupling strength, as well as their long coherence times and low dissipation, cavity magnonic systems hold significant potential in quantum metrology. For example, by conducting precise measurements of the cavity field within these systems, weak magnetic fields can be detected with a level of precision that approaches the Heisenberg limit \cite{PhysRevB.109.L041301}.

In this paper, we propose a quantum gyroscope scheme based on a cavity magnomechanical system. The proposed quantum gyroscope consists of two optical whispering gallery modes (WGMs) and a single magnon mode within a magnetic insulator yttrium iron garnet (YIG) sphere. Benefting from the rich magnetic nonlinearity of the YIG sphere, the interaction between the magnon mode and the two optical WGMs is intrinsically nonlinear.  By utilizing a two-mode squeezed coherent state, we achieve a surpassing of the SNL under ideal conditions. Furthermore, we discovered that under non-Markovian conditions, the system's performance in a dissipative environment is influenced by the parameters of the environmental spectrum. We analyzed the effects of various spectral density forms on the system's performance. With appropriate parameter settings, the precision of the quantum gyroscope remains stable over time. Moreover, coherent states are relatively easy to prepare experimentally, and the cavity magnomechanical system can effectively reduce the size of the device, thus the proposed quantum gyroscope is practical.

\section{THEORETICAL MODEL}

\begin{figure}
\centering
\includegraphics[width=2.5in]{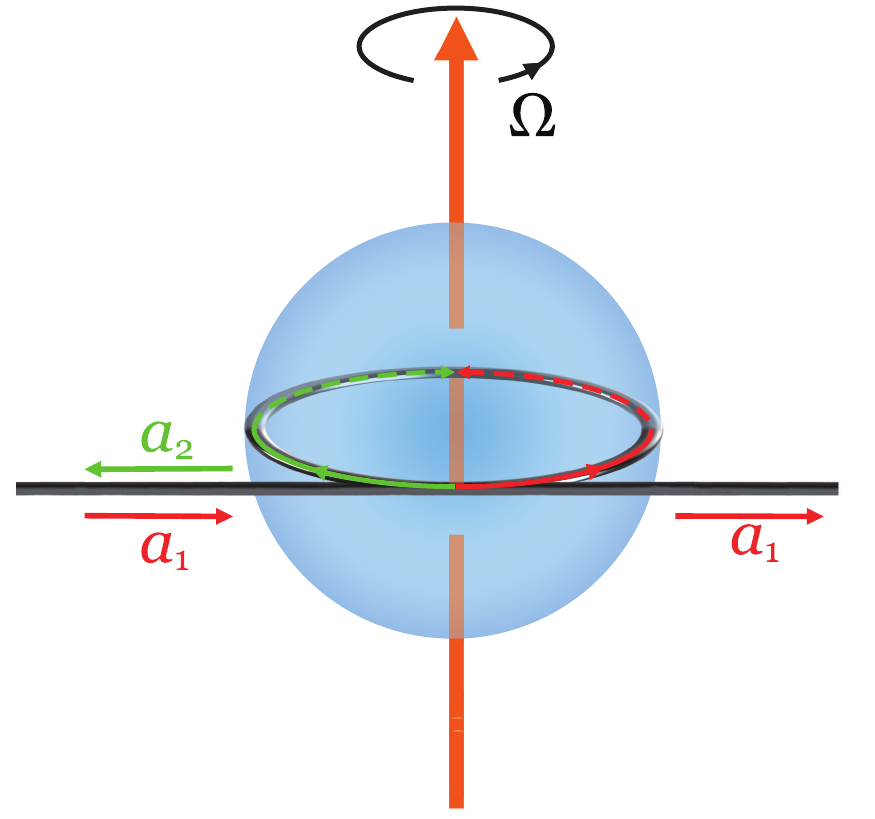}
 \caption{(Color online) Sketch of the system. The two optical WGMs are discriminated into transverse electric (TE) and transverse magnetic (TM) modes, as well as a magnetostatic mode supported by a YIG sphere. }
 \label{fig:fig1}
\end{figure}

Our proposal is composed of two optical
WGMs and a magnetostatic mode supported by a YIG sphere, as shown in Fig. \ref{fig:fig1}. When the cavity
spins along the clockwise direction, the resonance frequency of cavity modes in the
resonator undergoes a Fizeau shift is \cite{Maayani2018}
 \begin{eqnarray}
 \Delta_F = \pm\Omega\frac{nr\omega_0}{c}(1-\frac{1}{n^2}-\frac{\lambda dn}{n d\lambda}), \label{eq1}
\end{eqnarray}
Here $\omega_0$ is the resonance frequency of a nonspinning resonator.
The parameters $\Omega$, $n$, and $r$ represent the angular velocity of the
spinning resonator, the refractive index and the radius of the resonator,
respectively. While $c$ and $\lambda$ represent the speed of light and the
wavelength of the microwave photon in a vacuum, respectively. The last term
in Eq. (\ref{eq1}), which is characterized by the relativistic origin of the
Sagnac effect, describes the dispersion term and is usually negligible
because it is typically small. Here, the Fizeau shifts $\Delta_F > 0$ and $\Delta_F < 0$
indicate that the cavity is rotating in the clockwise aand counterclockwise
directions, respectively.

The total Hamiltonian of this system can be expressed in the form ($\hbar=1$)
\begin{eqnarray}
 H =H_0+H_{\rm{OM}}+H_{\rm{D}},
\end{eqnarray}
where
\begin{eqnarray}
H_0= \sum_{j=1,2}(\omega _{j}\pm \Delta_F)a_j^{\dagger}a_j+\omega _{m}m^{\dagger}m \label{eq2}
\end{eqnarray}
is the free Hamiltonian,
\begin{eqnarray}
H_{\rm{OM}}= g_{0}(a_1^{\dagger}a_2m_{}^{\dagger}+a_1a_2^{\dagger}m) \label{eq3}
\end{eqnarray}
describes a three-wave mixing process and is the standard optomagnonic
interaction Hamiltonian with a coupling strength of  $g_{0}$, which has
been experimentally demonstrated using Brillouin light scattering. According
to the Hamiltonian $H_{\rm{OM}}$, the annihilation of a magnon and a
photon in one cavity mode will create a photon in another cavity mode. Here,
\begin{eqnarray}
  H_{\rm{D}}=\sum_{j=1,2}\Omega_j\{a_j^{\dag} \exp[-i(\omega_l t+\phi_l)]+\text{H.c}\}, \label{eq4}
\end{eqnarray}
is the the laser driven term. Here $a_j$ $(a_j^{\dagger})$ are the annihilation
(creation) operator for the $j$th optical WGMs with the frequencies of a
nonspinning resonator are $\omega _{j}$. Here, we ensure that the two optical
WGMs are discriminated into TE and TM modes and confined close to the equator
of the YIG sphere. The parameter $m$ ($m^{\dagger}$) represents the
annihilation (creation) operator for the magnon mode with frequency
$\omega_{m}$. This frequency can be adjusted to match the splitting
of two WGMs by an external magnetic field, such as $\omega_{m}=|\omega_{1}-\omega_{2}|$.
The parameters $ \Omega_j$,  $\omega_{l}$, and $\phi_l$ are the
Rabi frequency, frequency and initial phase of the driven laser field, respectively.

To solve the Hamiltonian $H$, we firstly apply a unitary transformation
$V=\exp \left[ -i\omega_{l}(a_1^{\dagger}a_1+ a_2^{\dagger}a_2)t\right]$ to transform the Hamiltonian
$H$ to $ H_{1}=V^{\dagger}HV+i\hbar \frac{\partial V}{\partial t}V$. Thus, we obtain
a new Hamiltonian

\begin{eqnarray}
 H_{1}=&& \sum_{j=1,2}\Delta_j a_j^{\dagger}a_j +\omega_{m}m^{\dagger}m
+  g_{0}(a_1^{\dagger}a_2m_{}^{\dagger}+a_1a_2^{\dagger}m) \nonumber \\&&
+\Omega_1(a_1^{\dagger}+a_1)+\Omega_2(a_2^{\dagger}+a_2),\label{eq6}
\end{eqnarray}
where $\Delta _{j}=\omega _{j}\pm \Delta_F-\omega _{l}$ are the detunings
of the two cavity modes.

The quantum Langevin equations for the system operators
of the linearized Hamiltonian in Eq. (\ref{eq6}) are given by
\begin{eqnarray}
\dot{a_{1}} &=& -i\Delta_1a_{1}-ig_0a_2m^{\dagger}-\frac{\kappa_1}{2}a_1+\sqrt{\kappa_1} a_{1in}-i\Omega_1,\nonumber \\
\dot{a_{2}} &=& -i\Delta_2a_{2}-ig_0a_1m-\frac{\kappa_2}{2}a_2+\sqrt{\kappa_2}a_{2in}-i\Omega_2,\nonumber \\
\dot{m} &=& -i\omega_{m} m-ig_0a_1^{\dagger}a_2-\frac{\gamma}{2}m+\sqrt\gamma{m_{in}},
\label{eq7}
\end{eqnarray}
where $\kappa_1$ and $\kappa_2$ are the decay rates of the cavity modes,
$\gamma$ is the decay rate of magnon mode, $a_{1in}$, $a_{2in}$ and $m_{in}$ are the corresponding noise operators
with zero mean value.

To gain further get more insight into the dynamics of the
cavity magnomechanical system, we apply the standard linearization
process. We rewriting all the bosonic operators as $a_j=\alpha_j+\delta a_j$,
$m=m_s+\delta m$. Here, $\alpha_j$ and $m_s$ are the steady-state mean values of
three modes, while $\delta a_j$ and $\delta m$ are the corresponding
quantum fluctuations terms, respectively. For convenience, we will still use $a_j$ and $m$ to represent $\delta a_j$ and $\delta m$, respectively. The linearized Hamiltonian of
the system is written as

\begin{align}
\dot{a_{1}} =& -i\Delta_1 a_{1}-ig_0\alpha_{2} m^{\dag}-ig_0m_s^{*}a_{2} -\frac{\kappa_1}{2}a_{1}+ \sqrt{\kappa_1}a_{1in},\nonumber \\
\dot{a_{2}} =& -i\Delta_2 a_{2}-ig_0\alpha_{1} m-ig_0m_s a_{1} -\frac{\kappa_2}{2}a_{2}+\sqrt{\kappa_2}a_{2in},\nonumber \\
\dot{m} =& -i\omega_{m} m-ig_0(\alpha_2a_1^{\dag}+\alpha_1^{*}a_2+a_1^{\dag}a_2)-\frac{\gamma}{2}m \nonumber\\
&+\sqrt\gamma{m_{in}}.
\end{align}
Next, we turn to the interaction picture, by using the transformation
$a_j\longrightarrow a_j e^{-i\Delta_j t}$,
$m\longrightarrow m e^{-i\omega_{m} t}$, the above equation can be
rewritten as

\begin{align}
\dot{a_{1}} &= -ig_0\alpha_{2}m^{\dag}e^{i(\Delta_1+\omega_{m}) t} -\frac{\kappa_1}{2}a_{1}+ \sqrt{\kappa_1}a_{1in},\nonumber \\
\dot{m} &= -ig_0\alpha_2 a_1^{\dag}e^{i(\Delta_1+\omega_{m}) t}-\frac{\gamma}{2} m +\sqrt\gamma{m_{in}}.
\end{align}
Here, we have considered that only the cavity mode $a_2$ is driven by a
laser. Thus, following the adiabatic approximation, we find that the
fluctuation of cavity mode $a_2$ decouples from the magnon mode $m$.
Also, during the derivation process, the small terms have been omitted, such as
$g_0\alpha_1^{*}\alpha_2$, $g_0m_s^{*}a_2 $, $g_0 m_sa_1$ and so on.
Therefore, the effective Hamiltonian corresponding to the above Langevin equation is
\begin{eqnarray}
H_e=\xi a_{1}^{\dag} m^{\dag}+\xi^{*}a_{1}m,
\end{eqnarray}
where $\xi=g_0\alpha_2e^{i(\Delta_1+\omega_m) t}$.
The evolution operator of the system is
\begin{eqnarray}
U(t)= e^{-i(\xi a_{1}^{\dag} m^{\dag}+\xi^{*}a_{1}m)t},
\label{eq10}
\end{eqnarray}
which is a two-mode squeezing interaction between the cavity
mode $a_1$ and the magnon mode $m$, and the corresponding
squeeze operator is $S(G)=e^{G^{*}a_{1}m-G a_{1}^{\dag} m^{\dag}}$. Here, $G=i\xi t$ represents
the squeeze parameter, which can be significantly enhanced
by an intense driving field of $\alpha_2$. In the Sagnac interferometer, the input and the output modes
are related to each other by the linear transformation
$b_1=\sum_{k=1}^{2}s_{1k}a_1$ and $b_2=\sum_{k=1}^{2} s_{2k}m$,
with $a$ and $m$ represent the input modes and $b_j$ are the output modes.
And $s_{jk}$ is the scattering matrix represented in the following form,
\begin{eqnarray}
s_{jk}=\left(
          \begin{array}{cc}
            S_{11} & S_{12}  \\
            S_{21}  & S_{22}  \\
          \end{array}
        \right)
        =
         \left(
          \begin{array}{cc}
            \cos(\frac{\phi}{2}) & \sin(\frac{\phi}{2})  \\
            -\sin(\frac{\phi}{2})  & \cos(\frac{\phi}{2})  \\
          \end{array}
        \right). \label{eq11}
\end{eqnarray}
Here, $\phi$ represents the Sagnac phase shift caused by rotation, which is equal to
\begin{eqnarray}
\phi=\frac{4\pi R^2\Omega}{\lambda c},
\label{eq12}
\end{eqnarray}
where $\lambda$ is the wavelength, $c$ is the light velocity in vacuum, and $R$ is  the radius of the WGM cavity.

Suppose the input state of the system is a two-mode coherent state $|\psi_{in}\rangle=|\alpha\rangle |\beta\rangle$, where $\alpha$ and $\beta$ are the complex amplitudes of the initial two-mode coherent light fields. It should be emphasized that two-mode coherent states, as a crucial non-classical resource, exhibit distinctive physical characteristics and technological advantages. These include inherent benefits in classical scalability and experimental feasibility, as well as a natural superiority in continuous-variable encoding schemes and quantum-enhanced precision measurement protocols features that make them particularly valuable for advanced quantum technological applications. Then, under the interaction of the system, the output state becomes a two-mode squeezed coherent state $|\psi_{out}\rangle=S(G)|\alpha\rangle |\beta\rangle=|\alpha,\beta,G\rangle$.

To find out the Sagnac phase shift, we measure the intensity difference operator between the two output beams, that is
\begin{align}
n_{d}&= b_1^{\dagger} b_1 - b_2^{\dagger} b_2 \nonumber \\
&=(a_1^{\dagger} a_1-m^{\dagger} m)\cos\phi+(a_1^{\dagger} m+ a_1 m^{\dagger})\sin\phi.
\end{align}
The average number of the intensity difference operator is \cite{PhysRevA.102.022614}
\begin{align}
\langle n_{d}\rangle=& \cos\phi(|\alpha|^{2} - |\beta|^{2})+\sin\phi[\cosh(2G)(\alpha^{*}\beta+\alpha\beta^{*})  \nonumber \\
&-\frac{1}{2} \sinh(2G)(\alpha^{*^{2}}+\alpha^{2}+\beta^{*^{2}}+\beta^{2})].
\end{align}
Then the sensitivity of sensing $\phi$ can be evaluated via the error propagation formula as:
\begin{align}
\delta\phi&=\frac{\delta n_d}{|\partial_\phi \langle n_d \rangle|}  \nonumber \\
&=\frac{\sqrt{\cos^{2}\phi (|\alpha|^{2} + |\beta|^{2})+ A\sin(2\phi)+ B\sin^{2}\phi}}{|-\sin\phi(|\alpha|^{2} - |\beta|^{2})+C\cos\phi|}, \label{eq17}
\end{align}
where $A=\alpha^{2}-\beta^{2}+\alpha^{*^{2}}-\beta^{*^{2}}$,\\
$B=[(\sinh^{4}(G) + \cosh^{4}(G))(|\alpha|^{2} + |\beta|^{2})-\sinh(4G)(\alpha^{*} \beta^{*} + \alpha \beta)+\sinh^{2}(G)\cosh^{2}(G)(6|\alpha|^{2} + 6|\beta|^{2}+4)]$,\\
and $C=[\cosh(2G)(\alpha^{*}\beta+\alpha\beta^{*})-\frac{1}{2} \sinh(2G)(\alpha^{*^{2}}+\alpha^{2}+\beta^{*^{2}}+\beta^{2})]$.

Focusing on the full-cycle phase accumulation regime ($\phi=2n\pi$, $n \in \mathbb{Z}$) that corresponds to Sagnac interferometer's fundamental sensitivity limit, we specialize to the $n=1$ case. Thus, Eq. (\ref{eq17}) reduces to:
\begin{align}
\delta\phi&=\frac{\sqrt{|\alpha|^{2} + |\beta|^{2}}}{|D\cosh(2G)-\frac{1}{2} F\sinh(2G)|},
\end{align}
where $D=\alpha^{*}\beta+\alpha\beta^{*}$ and $F=\alpha^{*^{2}}+\alpha^{2}+\beta^{*^{2}}+\beta^{2}$. It is evident that the phase sensitivity in this configuration is systematically determined by three key parameters: the magnitude of two-mode squeezing  $G$, and the complex amplitude $\alpha$ and $\beta$ of the initial coherent state components. To establish a rigorous benchmark for evaluating quantum-enhanced phase estimation capabilities, we quantitatively compare the achieved sensitivity to the SNL: $\delta\phi_{\text{SNL}}=\frac{1}{\sqrt{\langle N \rangle}}$, where $\langle N \rangle$ represents the average number of the total photons. For the output state, the average photon number can be written as:
\begin{align}
\langle N \rangle =& \langle b_1^{\dagger} b_1 + b_2^{\dagger} b_2 \rangle = \langle a^{\dagger} a + m^{\dagger} m \rangle \nonumber \\
=& (|\alpha|^{2} + |\beta|^{2})\cosh(2G)- \sinh(2G)(\alpha^{*} \beta^{*} + \alpha \beta)  \nonumber \\
&+ 2 \sinh^{2}(G).\label{eq13}
\end{align}
Generally, the average photon numbers of the coherent state are higher than that of the squeezed state, i.e.,$|\alpha|^{2},|\beta|^{2}\gg \sinh^{2}(G)$. Therefore, Eq. (\ref{eq13}) can be approximated as:
\begin{eqnarray}
\langle N \rangle \approx (|\alpha|^{2} + |\beta|^{2})\cosh(2G)-(\alpha^{*}\beta^{*}+\alpha\beta)\sinh(2G), \nonumber\\
\end{eqnarray}
here, for simplicity, we assume that the squeezing parameter \textit{G} is a real number.

In the following section, three specific cases will be examined individually to establish the relationship between phase sensitivity and the SNL under each condition. In the following three cases, the parameters are expressed as $\alpha=|\alpha|e^{i\varphi_1}$ and $\beta=|\beta|e^{i\varphi_2}$, where $\varphi_1$ and $\varphi_2$ are the phases of the complex amplitudes $\alpha$ and $\beta$, respectively.

(i)$\alpha=\beta=|\alpha|e^{i\varphi}$

In this condition, the average number of total photons can be simplified as:
\begin{align}
\langle N \rangle &=2|\alpha|^2[\cosh(2G)-\sinh(2G)\cos2\varphi].
\end{align}
Meanwhile, the corresponding uncertainty of the phase shift is
\begin{align}
\delta\phi&=\frac{\sqrt{2|\alpha|^{2}}}{2|\alpha|^2\cosh(2G)-2|\alpha|^2\sinh(2G)\cos2\varphi}\nonumber \\
&=\frac{1}{k}\frac{1}{\sqrt{\langle N \rangle}}.
\end{align}
Here, $\frac{1}{k}$ represents the ratio of the phase shift uncertainty to the SNL, and it can be expressed as:
\begin{align}
\frac{1}{k}&=\frac{1}{[\cosh(2G)-\sinh(2G)\cos2\varphi]^{\frac{1}{2}}}.
\end{align}
Since $\cos2\varphi \in [-1,1]$, it follows that $\sinh(2G)\cos2\varphi \in [-\sinh(2G),\sinh(2G)]$. Note that $\cosh(2G)>\sinh(2G)$, which ensures that $k>0$ and $k$ is a real number. It is easy to see that $\frac{1}{k}$ reaches its maximum value when $\cos2\varphi=1$ and its minimum value when $\cos2\varphi=-1$.
\begin{align}
\max(\frac{1}{k})&=\frac{1}{[\cosh(2G)-\sinh(2G)]^{\frac{1}{2}}}=e^{G}, \\
\min(\frac{1}{k})&=\frac{1}{[\cosh(2G)+\sinh(2G)]^{\frac{1}{2}}}=e^{-G}. \label{eq23}
\end{align}
Since $e^{-G}<1$ , it can be seen from Eq. (\ref{eq23}) that, when measuring the phase shift, the gyroscope under consideration exhibits a reduced uncertainty in comparison to a classical gyroscope. This implies that the uncertainty in measuring the angular velocity $\Omega$ is also correspondingly reduced. Moreover, as the squeezing parameter \textit{G} increases, the uncertainty decreases exponentially. Therefore, we can enhance the sensitivity of  gyroscopes by adjusting the phase of the initial coherent state, specifically setting $\varphi=\frac{(2n+1)\pi}{2}$, $n=0,1,2,\dots$, and appropriately increasing the squeezing parameter \textit{G}.

(ii)$|\alpha|=|\beta|,\varphi_1\ne \varphi_2$

In this case, $\langle N \rangle$ can be written as:
\begin{align}
\langle N \rangle &=2|\alpha|^2[\cosh(2G)-\sinh(2G)\cos(\varphi_1+\varphi_2)].
\end{align}
Meanwhile, we can calculate the uncertainty of the phase shift as:
\begin{align}
\delta\phi&=\frac{1}{\sqrt{\langle N \rangle}|\cos\Delta\varphi|[\cosh(2G)-\sinh(2G)\cos\varphi_p]^{\frac{1}{2}}}\nonumber \\
&=\frac{1}{k}\frac{1}{\sqrt{\langle N \rangle}}.
\end{align}
\begin{figure}[ht]
	\centering

	\begin{minipage}[t]{0.5\columnwidth}
		\centering
		\includegraphics[width=\textwidth]{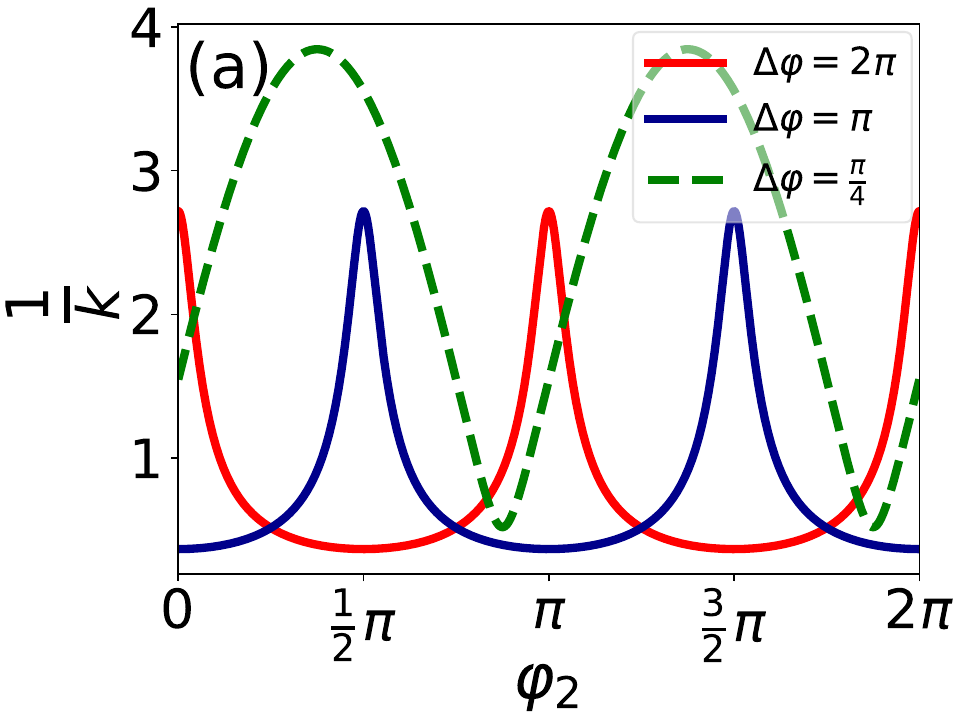}

		\label{fig:2a}
	\end{minipage}%
	\hfill
	\begin{minipage}[t]{0.5\columnwidth}
		\centering
		\includegraphics[width=\textwidth]{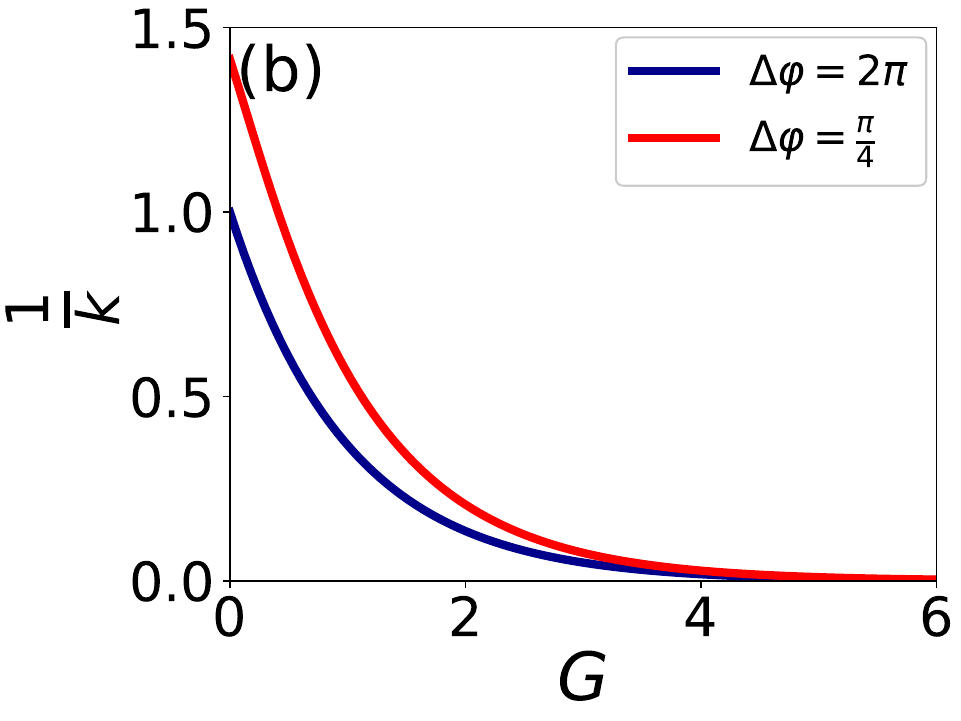}

		\label{fig:2b}
	\end{minipage}

	\vspace{1em}
    \caption{For case II, the value of $\frac{1}{k}$ depends on $\varphi_2$ and the squeezing parameter $G$. (a)$\frac{1}{k}$ as a function of $\varphi_2$, here $G=1$. (b)$\frac{1}{k}$ as a function of $G$, showing an exponential decay, here $\varphi_2=\frac{\pi}{2}$.}
    \label{fig:fig2}
\end{figure}
Here,
\begin{align}
\frac{1}{k}&=\frac{1}{|\cos\Delta\varphi|[\cosh(2G)-\sinh(2G)\cos\varphi_p]^{\frac{1}{2}}},
\end{align}
where $\Delta\varphi=\varphi_1-\varphi_2$, $\varphi_p=\varphi_1+\varphi_2$.
Obviously, the value of $\frac{1}{k}$ is related to the phase difference $\Delta\varphi$. Therefore, it is essential to provide a systematic discussion of this phase difference. First, we consider a special case, where $\Delta\varphi=\frac{\pi}{2}+n\pi$($n=1,2,3\dots$). This will lead to $\frac{1}{k} \to \infty$. To understand why this situation occurs, we need to revisit the expression for $\langle n_d \rangle$:
\begin{align}
\langle n_{d}\rangle=&\sin\phi\cos\Delta\varphi[\cosh(2G)-\sinh(2G)\cos\varphi_p]\nonumber\\
=&0.
\end{align}
This implies that, due to the interference between the two coherent states, the information about the phase shift carried by the intensity difference operator is completely lost.

Next, we consider the case where $\Delta\varphi=n\pi$, i.e., $|\cos\Delta\varphi|=1$. In this case, $\frac{1}{k}$ can be written as:
\begin{align}
\frac{1}{k}=\frac{1}{[\cosh(2G)-\sinh(2G)\cos(2\varphi_2+n\pi)]^{\frac{1}{2}}}.
\end{align}
\begin{figure}[ht]
	\centering

	\begin{minipage}[t]{0.5\columnwidth}
		\centering
		\includegraphics[width=\textwidth]{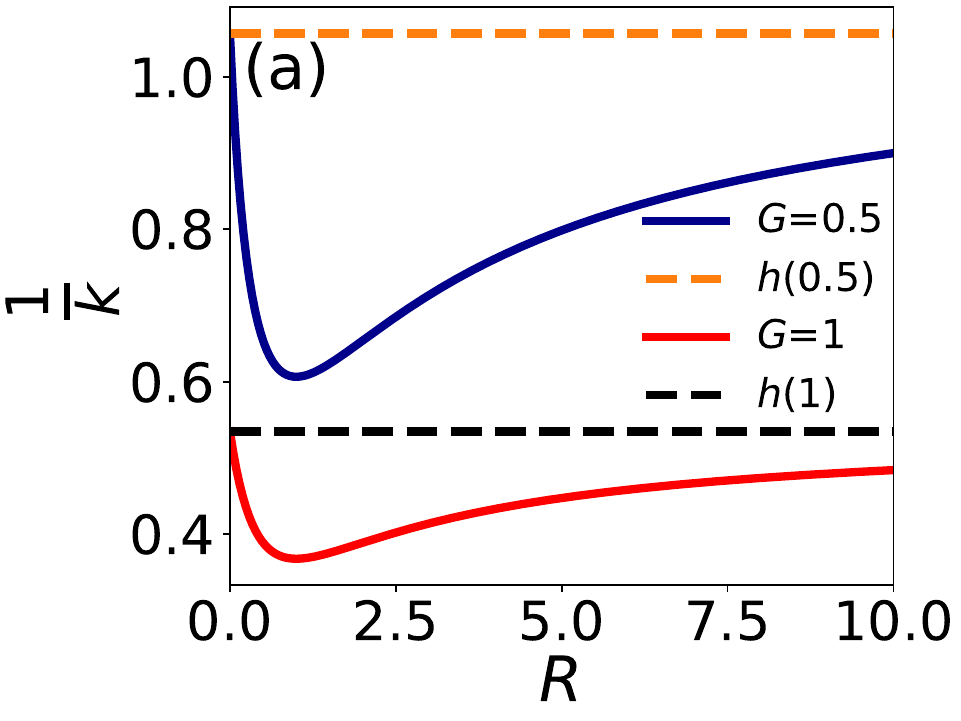}

		\label{fig:2a}
	\end{minipage}%
	\hfill
	\begin{minipage}[t]{0.5\columnwidth}
		\centering
		\includegraphics[width=\textwidth]{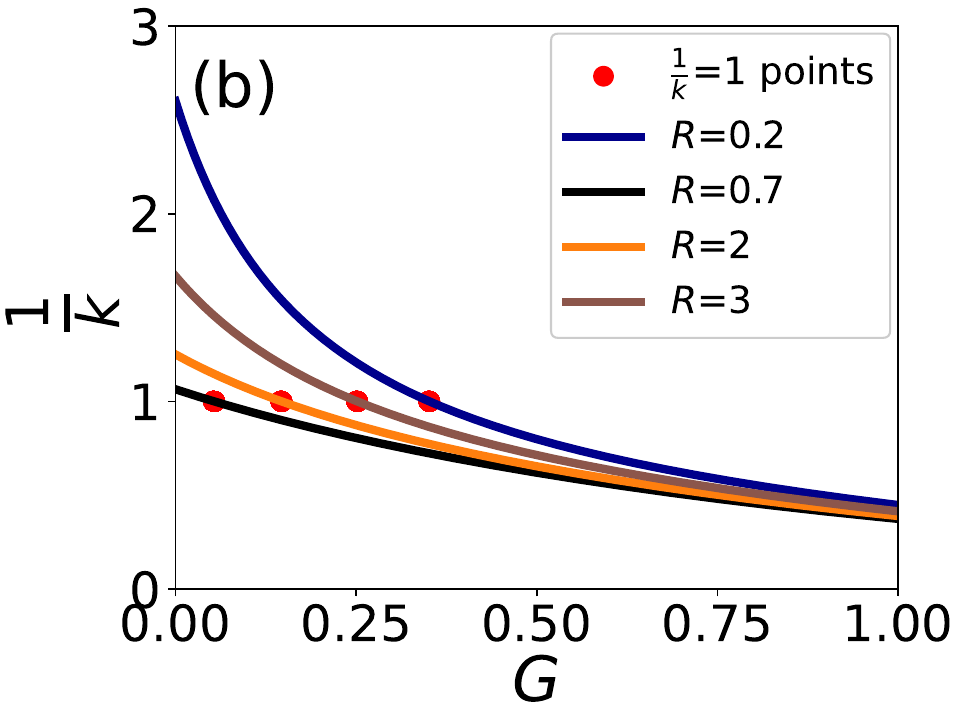}

		\label{fig:2b}
	\end{minipage}

	\vspace{1em}
    \caption{ For case III, the value of $\frac{1}{k}$ depends on $R$, and the squeezing parameter $G$. (a)$\frac{1}{k}$ as a function of $R$. The maximum value of $\frac{1}{k}$ (corresponding to the limits $R \to 0$ or $R \to \infty$) is indicated by orange and black dashed lines for $G=0.5$ and $G=1$, respectively. (b)$\frac{1}{k}$ as a function of $G$. The parameter is $\varphi=\frac{\pi}{2}$.}
    \label{fig:fig3}
\end{figure}
In order to gain further insight into the relationship between $\frac{1}{k}$ and the other parameters, we conduct a numerical simulation. In Fig. \ref{fig:fig2}(a), we plot the relationship between $\frac{1}{k}$ and $\varphi_2$ for a specific value of $\Delta\varphi$. As shown in the figure, $\frac{1}{k}$ is a periodic function with a period of $\varphi_2$. For the condition $\Delta\varphi=n\pi$, the maximum value, denoted as $e^{G}$ occurs at $\varphi_2=(k-\frac{n}{2})\pi$, where $k=0,1,2,\dots$, while the minimum value $e^{-G}$ is found at $\varphi_2=(k-\frac{n}{2})\pi+\frac{\pi}{2}$. Furthermore, as depicted by the green dashed line in the figure, the gyroscope's precision is reduced when $\Delta\varphi \ne n\pi$. This outcome is readily comprehensible. For $\Delta\varphi=n\pi$, the information transfer between the two coherent states fulfills the constructive interference condition, enabling maximal information transmission. Fig. \ref{fig:fig2}(b) illustrates the relationship between the minimum value of $\frac{1}{k}$ and the squeezing parameter \textit{G}. It is evident that for $\Delta\varphi=2\pi$, the decay of $\frac{1}{k}$ with increasing $G$ is accelerated, aligning with our preceding analysis. It is important to note that, as shown in Fig. \ref{fig:fig2}, $\frac{1}{k}$ is not always less than 0. Its value depends on the specific choices of $\varphi_1$, $\varphi_2$, and \textit{G}.

(iii)$|\alpha|\ne|\beta|,\varphi_1=\varphi_2=\varphi$

In this case, the parameter $\langle N \rangle$ is
\begin{align}
\langle N \rangle &=(|\alpha|^2+|\beta|^2)\cosh(2G)-2\sinh(2G)|\alpha||\beta|\cos2\varphi ,
\end{align}
and $\delta\phi$:
\begin{align}
\delta\phi&=\frac{\sqrt{|\alpha|^{2} + |\beta|^{2}}}{|2|\alpha||\beta|\cosh(2G)-\sinh(2G)\cos2\varphi(|\alpha|^{2} + |\beta|^{2})|}.  \nonumber \\
\end{align}
Setting $\frac{|\beta|}{|\alpha|}= R$, the corresponding $\frac{1}{k}$ is:
\begin{align}
\frac{1}{k}&=\frac{(1+R^2)^{\frac{1}{2}}[(1+R^2)\cosh(2G)-2R\sinh(2G)\cos2\varphi]^{\frac{1}{2}}}{|2R\cosh(2G)-
\sinh(2G)(1+R^2)\cos2\varphi|}.\label{eq30}
\end{align}
Partial differentiation of $\frac{1}{k}$ with respect to $\varphi$ indicates a local minimum of $\frac{1}{k}$ at $\varphi=\frac{\pi}{2}+n\pi$. Consequently, we proceed to consider the case of $\varphi=\frac{\pi}{2}$. First, we will investigate the dependence of $\frac{1}{k}$ on $R$. Defining
\begin{align}
f(R)=\frac{(1+R^2)^{\frac{1}{2}}[(1+R^2)\cosh(2G)+2R\sinh(2G)]^{\frac{1}{2}}}{|2R\cosh(2G)+\sinh(2G)(1+R^2)|},
\end{align}
differentiation with respect to $R$ reveals that $f(R)$ is monotonically decreasing for $R \in [0,1)$ and monotonically increasing for $R \in [1,\infty)$, as shown in Fig. \ref{fig:fig3}(a). Thus, $f(R)$ exhibits a local minimum of $e^{-G}$ at $R = 1$, which aligns with case I. Notably, for $R \to 0$ or $R \to \infty$, $\frac{1}{k} \to \frac{\sqrt{\cosh(2G)}}{\sinh(2G)}$. We define $h(G)=\frac{\sqrt{\cosh(2G)}}{\sinh(2G)}$, and its derivative is:
\begin{align}
h^{'}(G)&=\frac{\cosh^{-\frac{1}{2}}(2G)(\sinh^2(2G)-2\cosh^2(2G))}{\sinh^2(2G)}<0,
\end{align}
which means that $h(G)$ is a monotonically decreasing function, and $h(G)\in (0,\infty)$. For $G_0=\frac{1}{2}\ln\frac{1+\sqrt{5}+\sqrt{2+2\sqrt{5}}}{2}$, $h(G_0)=1$. This implies that when $G>G_0$, $\frac{1}{k}<1$ always holds, as clearly illustrated by the red solid line Fig. \ref{fig:fig3}(a). Significantly, an increase in $G$ leads to a decrease in $h(G)$, consequently resulting in a downward shift of the entire function curve. Therefore, for each $R$ value, there exists a corresponding $G_{th}$, such that when $G>G_{th}$ the condition $\frac{1}{k}<1$ holds for that $R$ value, as depicted in Fig. \ref{fig:fig3}(b). The $G_{th}$ values are given by the x-coordinates of the red dots in the figure. From the monotonic behavior of $f(R)$, it follows that in the range $R \in [0,1)$, $G_{th}$ exhibits a decreasing trend with increasing $R$, whereas for $R \in [1,\infty)$, $G_{th}$ shows an increasing trend with increasing $R$. In summary, achieving $\frac{1}{k}<1$ is dependent on both the $R$ value and its corresponding $G_{th}$.

From the preceding analysis of three specific cases, we conclude that the proposed gyroscope scheme attains its maximum precision when operating under the conditions of the first case, namely $\alpha=\beta$. The enhanced clarity and strength of the rotation signal when $\alpha=\beta$ arises from the interference mechanism between the two beams. This can be understood by analogy with the interference of two classical optical beams, where the interference fringes are characterized by optimal clarity and highest contrast when the beam amplitudes are identical.

\section{EFFECTS OF DISSIPATIVE ENVIRONMENTS}

The decoherence effect in quantum systems presents a
significant challenge for the practical application of
quantum sensors. In our quantum gyroscope, the primary
source of decoherence is photon dissipation caused
by energy exchange with the environment. Generally,
Markovian or non-Markovian approximations are employed
to describe the interaction between quantum systems
and the environment. In the Markovian approximation,
the perturbative method, primarily in the form of a
generalized Lindblad master equation, serves as the main approach for modeling the influence of the environment.
While the non-Markovian approximation will  induce
diverse characteristics that are absent in the Markovian approximation \cite{PhysRevLett.109.170402,PhysRevLett.121.220403,RevModPhys.88.021002}.
Therefore, the characterization of the environment
becomes particularly important. To investigate
the impact of photon dissipation on our quantum
gyroscope, we utilize a discretization of a
continuous environment. We assume that photon dissipation arises from the energy exchange between
the two fields and two independent bosonic environments.
The Hamiltonian of the total system is:
\begin{eqnarray}
 H =H_0+\sum\limits_{k=a,m}\sum\limits_{l}[\omega_{k,l}b_{k,l}^{\dag}b_{k,l}+g_{k,l}(k^{\dag}b_{k,l}+ \mathrm{H.c.})], \nonumber \\
 \end{eqnarray}
where $b_{k,l}$ are annihilation operator of the $l$th
mode with frequency $\omega_{k,l}$ of the environment, and
$g_{k,l}$ is the coupling strength between the caviy (magnon) mode and
the corresponding environment and is characterized by
the spectral density function, $J_k(\omega)=\sum\limits_{l}g_{k,l}^2 \delta(\omega-\omega_l)$.
The environments are consider as the Ohmic-family
spectral density in the from of \cite{RevModPhys.59.1}
\begin{eqnarray}
J_1(\omega)=J_2(\omega)=\gamma\omega^s \omega_c^{1-s} e^{-\frac{\omega}{\omega_c}},
 \end{eqnarray}
where $\gamma$ is the coupling strength, $\omega_c$ is the cut-off frequency, and $s$ is the Ohmicity index.

From Eq. (\ref{eq11}), we can derive:
\begin{align}
b_1&=\cos\frac{\phi}{2}a_1+\sin\frac{\phi}{2}m=e^{-i\phi U_2}ae^{i\phi U_2},  \nonumber\\
b_2&=-\sin\frac{\phi}{2}a_1+\cos\frac{\phi}{2}m=e^{-i\phi U_2}me^{i\phi U_2},
\end{align}
where $U_2=\frac{1}{2i}(a_1^{\dagger}m-a_1 m^{\dagger})$. And $U_2$ can be expressed as~\cite{Wang:18}:
\begin{align}
U_2&=VU_{\phi}V \nonumber\\
&=e^{i\frac{\pi}{4}(a_1^{\dagger}m+a_1 m^{\dagger})}e^{-i\frac{\phi}{2}(a_1^{\dagger}a_1-m^{\dagger}m)}e^{i\frac{\pi}{4}(a_1^{\dagger}m+a_1 m^{\dagger})}.
\end{align}
Then, the entire system can be viewed as the input passing through a 50-50 beam splitter, undergoing interaction, and then passing through another 50-50 beam splitter before being output. Thus, we can equivalently write $H_0$ as:
\begin{align}
H_0&=\omega_0a_1^{\dagger}a_1+\omega_0m^{\dagger}m+\Omega(a_1^{\dagger}a_1-m^{\dagger}m),
\end{align}
where $\phi=2\Omega t$. Next, we can derive an exact master equation for the encoding process using the Feynman-Vernon influence functional method \cite{PhysRevA.76.042127,kam2023coherent},
\begin{align}
\dot{\rho}(t) &= \sum\limits_{k=a_1,m} \left\{
    -i\varDelta_k(t)\big[k^{\dagger}k, \rho(t)\big]
    + \varGamma_k(t)\mathcal{D}_k\rho(t)
\right\}, \label{eq37}
\end{align}
where $\mathcal{D}_k\rho(t)=2k\rho k^{\dagger}-\left\{k^{\dagger}k,\rho\right\}$, $\varDelta_k(t)=-\rm{Im}[\frac{\dot{\textit{u}}_\textit{k}(t)}{\textit{u}_\textit{k}(t)}]$, and $\varGamma_k(t)=-\rm{Re}[\frac{\dot{\textit{u}}_\textit{k}(t)}{\textit{u}_\textit{k}(t)}]$. The effects related to the environment are all encapsulated in the function $u_k(t)$, and $u_k(t)$ satisfies \cite{Jiao:23}
\begin{align}
\dot{u}_k(t)+i\omega _k u_k(t)+\int_0^t f(t-\tau)u_k(\tau) \, d\tau&=0 ,\label{eq38}
\end{align}
where $\omega _{a_1,m}=\omega_0\pm\Omega$, and $f(x)=\int_0^\infty J(\omega)e^{-i\omega x} \, d\omega$ is the environmental correlation function. For convenience in writing the following expressions, we set $u_{a_1}(t)=u_1(t)$, and $u_m(t)=u_2(t)$. Solving Eq. (\ref{eq37}), we obtain
\begin{widetext}
\begin{align}
\langle n_{d} \rangle=\frac{ix(B_1-B_2)\exp[\frac{m_1q_1^2+\bar{m}_1n_1^2+n_1q_1}{(p_1-1)^2-4|m_1|^2}+\frac{m_2q_2^2+\bar{m}_2n_2^2+n_2q_2}{(p_2-2)^2-4|m_2|^2}+e_1+e_2]}{[(p_2-1)^2-4|m_2|^2]^{\frac{1}{2}}[(p_2-1)^2-4|m_2|^2]^{\frac{1}{2}}} \label{eq39}
\end{align}
\end{widetext}
where $x=(\sqrt{A_1A_2}\cosh^2 G)^{-1}$, $m_l=\frac{-iu_l(t)^2\tanh G}{2A_l}$, $p_l=|u_l(t)|^2(1-A_l)^{-1}$, $A_l=1-[|u_l(t)|^2-1]^2\tanh^2 G$, $l=1,2$. Other parameters are too complex, so their derivations and forms are provided in Appendix.
For the sake of simplicity, we here set $\alpha$ = $\beta$.

Firstly, we consider the Born-Markovian approximation, which requires a weakly coupled probe-environment system and a characteristic timescale of $f(t-\tau)$ that is substantially shorter than the probe's intrinsic dynamics. Their approximate solutions read $u_l(t)=e^{[-\kappa_l+i(\omega_l+\Delta)]t}$, with $\kappa_l=\pi J(\omega_l)$ and $\Delta=\mathcal{P}\int_{0}^\infty  \frac{J(\omega)}{\omega_l-\omega} \, d\omega$. Under ideal conditions, $u_l(t)$ can be further simplified to $u_l(t)=e^{-i\omega_l t}$.
More generally, the non-Markovian case should be taken into account. In this case, the solution for $u_l(t)$ depends on the following two expressions:
\begin{align}
&\omega_l - \int_0^\infty \frac{J(\omega)}{\omega-E_l} \, d\omega=E_l,  \label{eq40}\\
&u_l(t)=Z_le^{-iE_{b,l} t}+\int_0^\infty \Theta(\omega)e^{-i\omega t} \, d\omega, \label{eq41}
\end{align}
where $Z_l=[1+\int_0^\infty \frac{J(\omega)}{(\omega-E_l)^2} \, d\omega]^{-1}$, and $\Theta(\omega)=\frac{J(\omega)}{[\omega-\omega_l-\Delta(\omega)]^2+[\pi J(\omega)]^2}$.  It is particularly noteworthy that $E_{b,l}$ represents the solution of Eq. (\ref{eq40}) when $E_l<0$. Eq. (\ref{eq41}) provides the form of the solution for $u_l(t)$, and it can be seen that it depends on the solution $E_l$ of Eq. (\ref{eq40}). So let us first study the properties of the solution to Eq. (\ref{eq40}).

We define the term on the left-hand side of Eq. (\ref{eq40}) as $P_l(E_l)$. When $E_l>0$, The function $P_l(E_l)$ has poles along the integration path, and they have infinite roots in this regime. If $E_l<0$, the equation has no solution, i.e., $E_{b,l}$ does not exist, then as $t \to \infty$, $u_l(t) \to 0$. This is because the solution for $u_l$ depends on its integral term, which approaches 0 in the long-time limit. From another perspective, $\varGamma_k(t)$ in Eq. (\ref{eq37}) represents the dissipation rate, and it does not equal to zero in the long-time limit.
\begin{figure}[ht]
    \centering
    \begin{minipage}[t]{0.48\columnwidth}
        \centering
        \includegraphics[width=\textwidth]{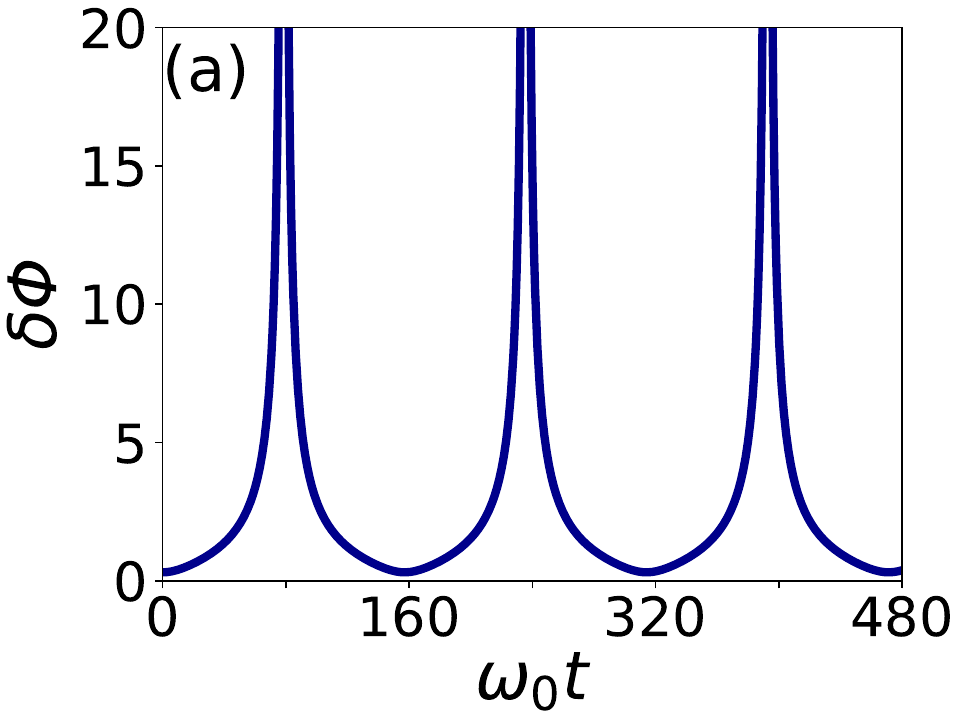}

        \label{fig:2a}
    \end{minipage}\hfill
    \begin{minipage}[t]{0.48\columnwidth}
        \centering
        \includegraphics[width=\textwidth]{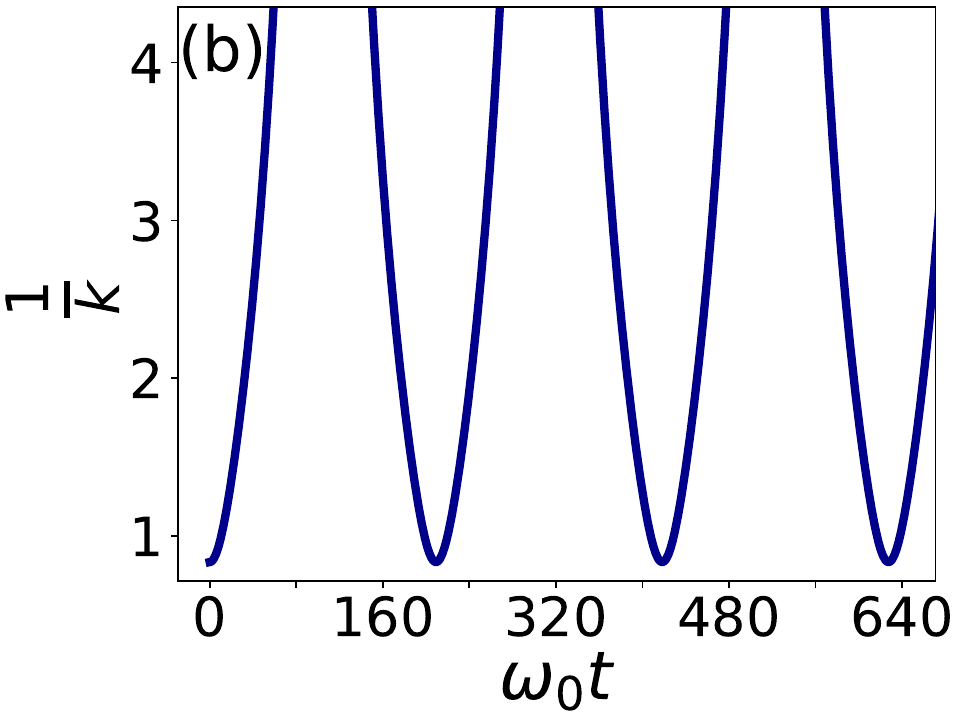}

        \label{fig:2b}
    \end{minipage}

    \vspace{1em}

    \begin{minipage}[t]{0.48\columnwidth}
        \centering
        \includegraphics[width=\textwidth]{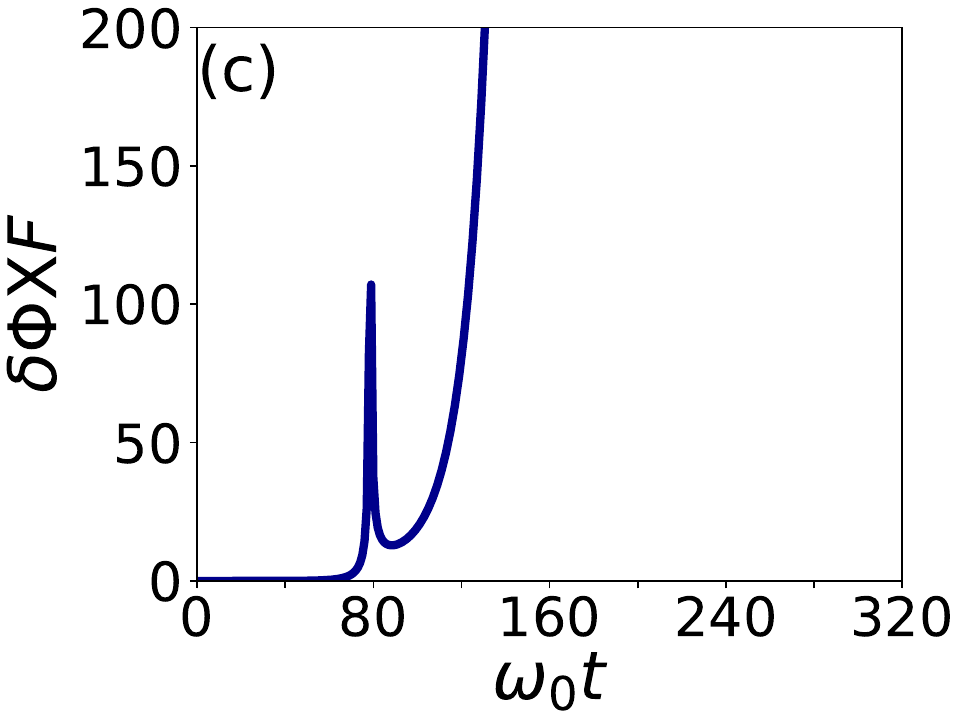}

        \label{fig:2c}
    \end{minipage}\hfill
    \begin{minipage}[t]{0.48\columnwidth}
        \centering
        \includegraphics[width=\textwidth]{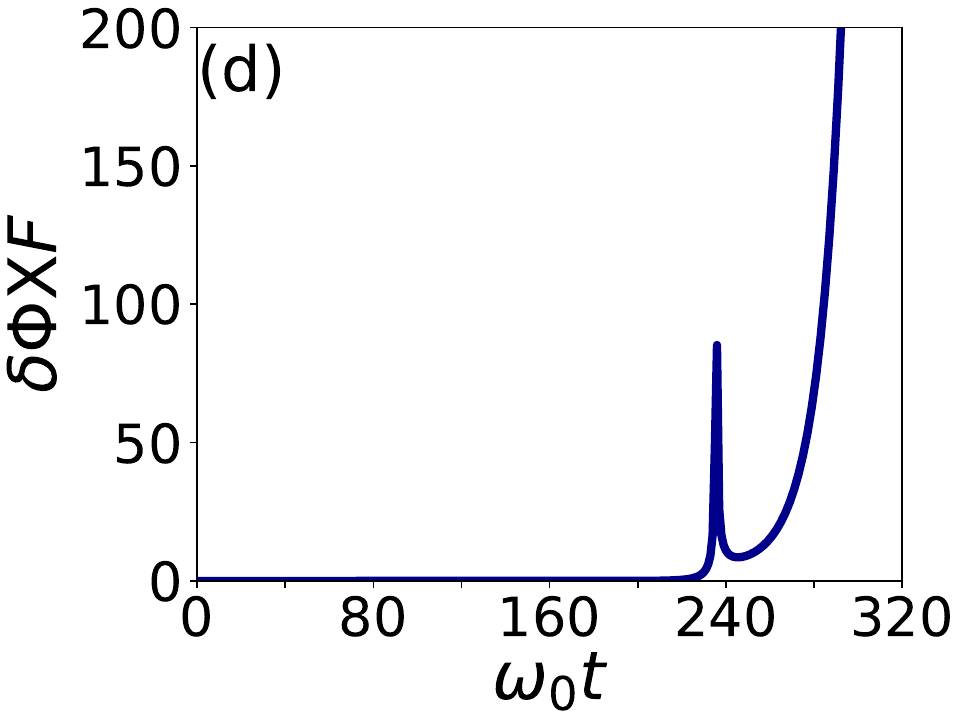}

        \label{fig:2d}
    \end{minipage}

     \caption{(a) The evolution of $\delta\phi$ over time in the ideal case; the global minimum is marked by the red dot. (b) The evolution of $1/k$ over time in the ideal case.
    (c) The dynamical evolution of $\delta\phi$ under the Born-Markovian approximation with $F=10^{-3}$. (d) The dynamical evolution of $\delta\phi$ under the Born-Markovian approximation with $F=10^{-10}$.}
    \label{fig:fig4}
\end{figure}
\begin{figure*}[ht]
    \centering
    \begin{minipage}[t]{0.32\textwidth}
        \centering
        \includegraphics[width=\textwidth]{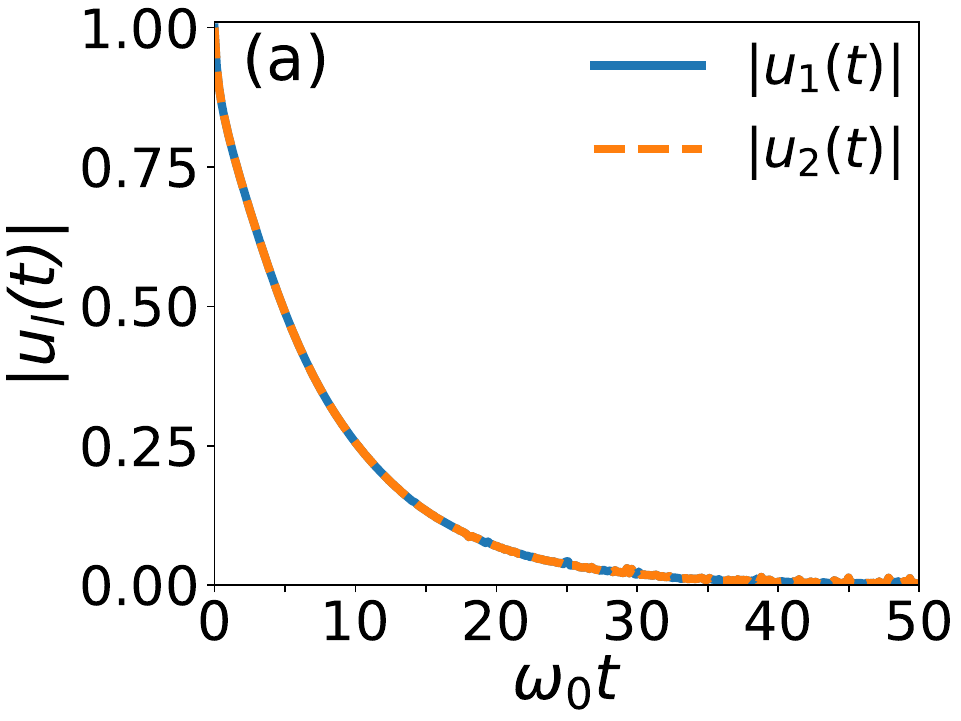}

        \label{fig:subfig_a}
    \end{minipage}\hfill
    \begin{minipage}[t]{0.32\textwidth}
        \centering
        \includegraphics[width=\textwidth]{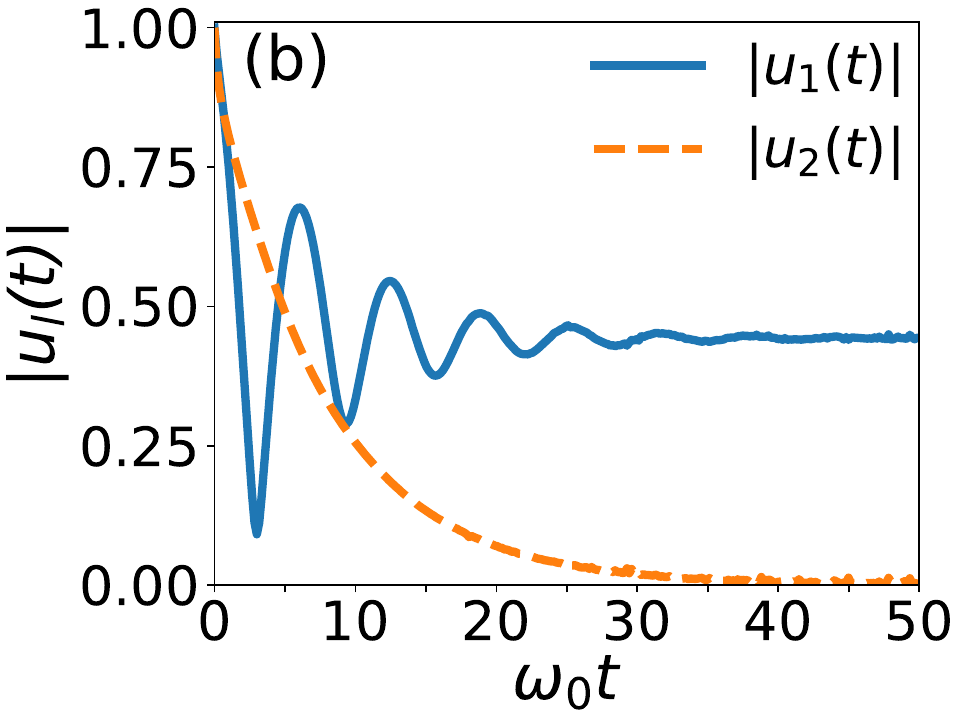}

        \label{fig:subfig_b}
    \end{minipage}\hfill
    \begin{minipage}[t]{0.32\textwidth}
        \centering
        \includegraphics[width=\textwidth]{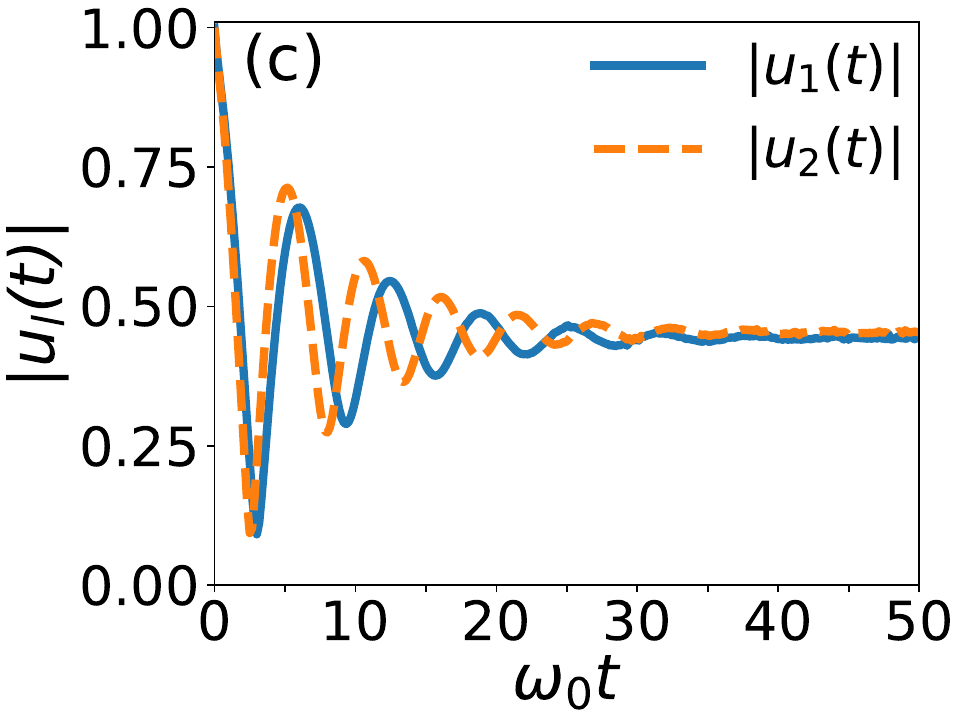}

        \label{fig:subfig_c}
    \end{minipage}

    \vspace{3em}

    \begin{minipage}[t]{0.32\textwidth}
        \centering
        \includegraphics[width=\textwidth]{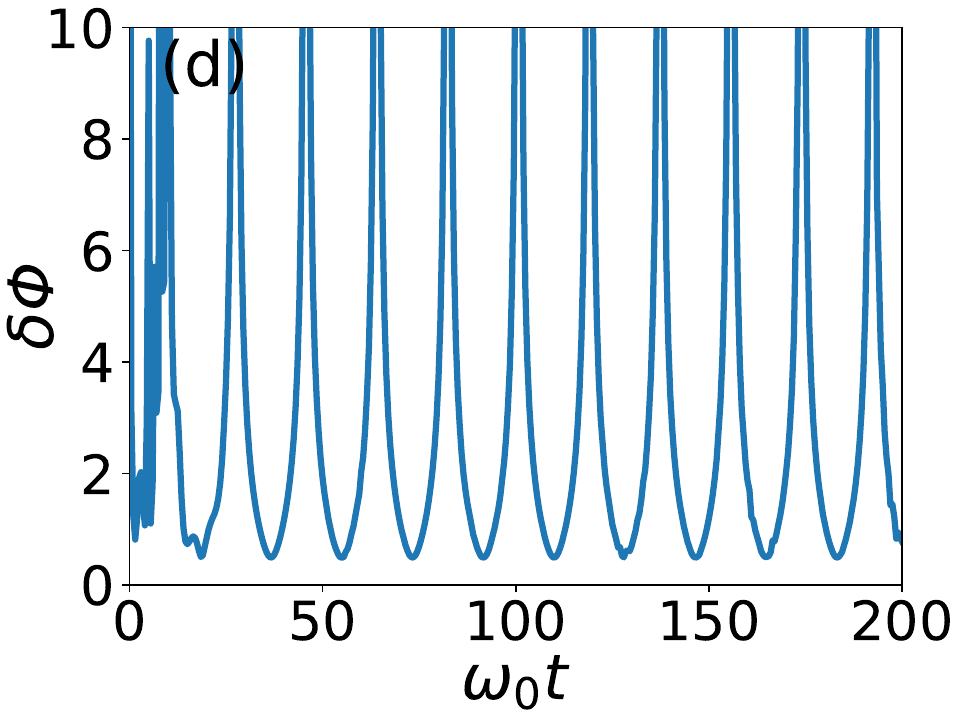} %

        \label{fig:subfig_d}
    \end{minipage}\hfill
    \begin{minipage}[t]{0.32\textwidth}
        \centering
        \includegraphics[width=\textwidth]{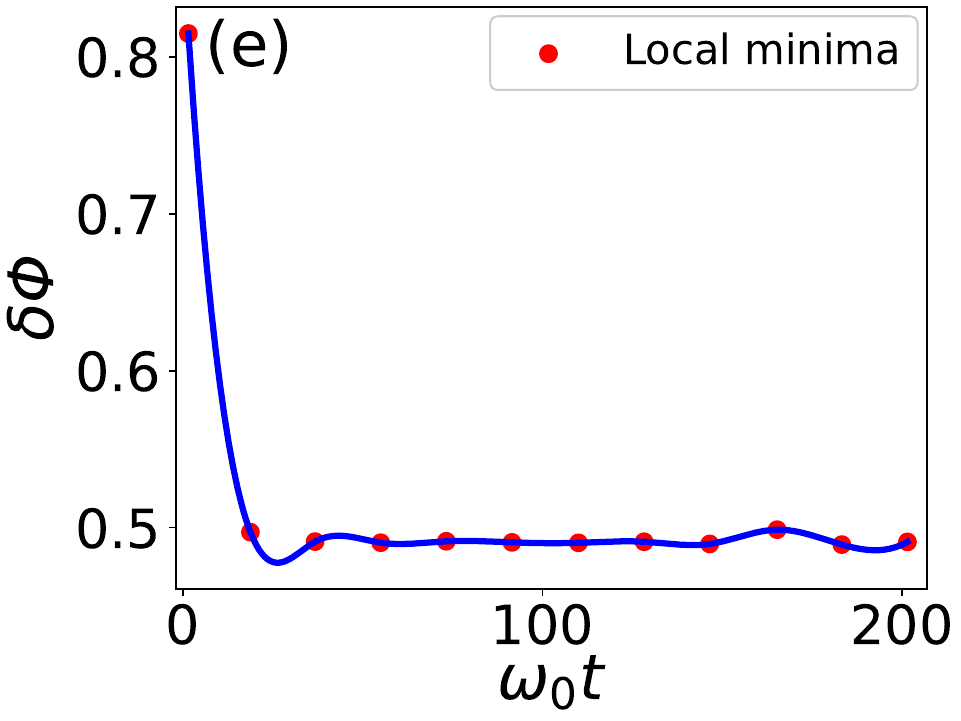}%
        \vspace{-0.5pt}

        \label{fig:subfig_e}
    \end{minipage}\hfill
    \begin{minipage}[t]{0.32\textwidth}
        \centering
        \includegraphics[width=\textwidth]{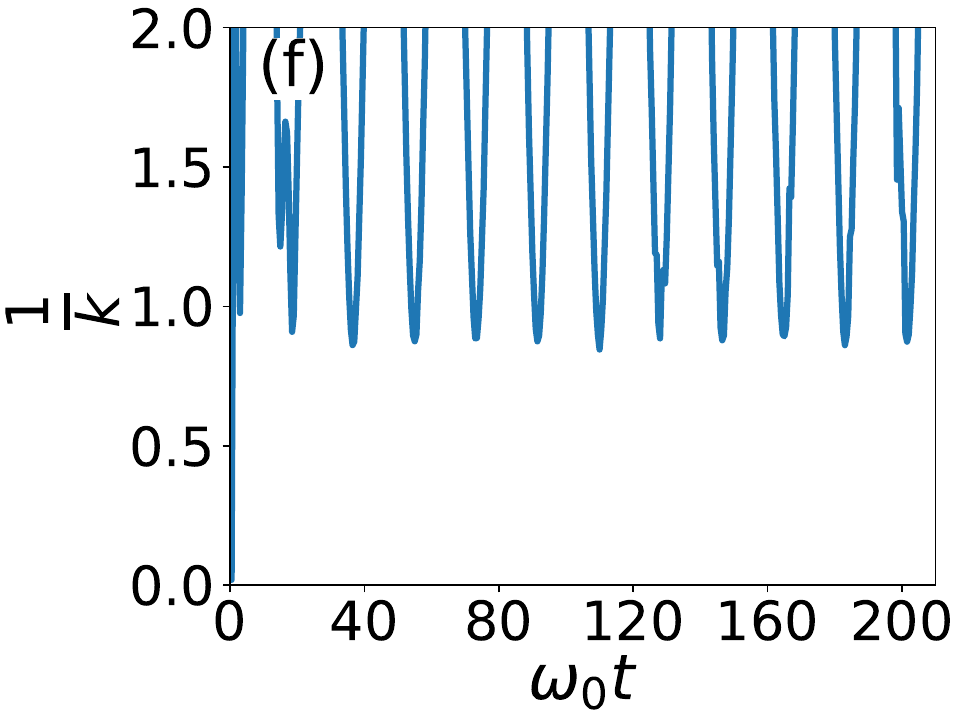} %

        \label{fig:subfig_f}
    \end{minipage}
    \caption{(a)-(c) Solutions for $u_l(t)$ at varying $\omega_c/\omega_0$: (a) $\omega_c/\omega_0=10$, (b) 20, and (c) 25.
    (d) The evolution of $\delta\phi$ with time under the condition $\omega_c=25\omega_0$. (e) The trajectory of the local minima of $\delta\phi$. (f) The evolution of $1/k$ over time. The parameters are $s=1$, $\gamma=0.05$, and $\Omega=0.1\omega_0$.}
    \label{fig:fig5}
\end{figure*}
At this point, although the system and the environment continuously exchange information, the information will eventually be completely leaked into the environment. This is precisely the situation we aim to avoid. If $E_l<0$ and Eq. (\ref{eq40}) has a solution $E_{b,l}$, then as $t \to \infty$, $u_l(t) \to Z_l e^{-iE_{b,l}t}$, which is referred to as bound states \cite{Jiao:23}. This indicates that, in the long-time limit, the value of $u_l(t)$ will oscillate around $Z_l$. This is because $\varGamma_k(t)$ will vanish in the steady-state limit, resulting in dissipationless oscillations. Consequently, there is always an interaction between the system and the environment, allowing the environment to return the information leaked by the system back to it. From the analysis above, it is clear that the behavior of $u_l(t)$ depends on whether Eq. (\ref{eq40}) has a solution when $E_l<0$. Through calculation, it is found that when the Ohmic-family spectral density satisfies the condition \cite{PhysRevLett.109.170402}
\begin{align}
&\gamma\omega_c\varGamma (s)>\omega_l, \label{eq42}
\end{align}
a bound state is formed, where $\gamma$ is the coupling strength, and $\varGamma (s)$ is the gamma function.

\section{NUMERICAL SIMULATIONS}
\hypertarget{sec:IV}{}
In this section, we demonstrate the effectiveness of our analysis through numerical simulations. We assume the initial two-mode coherent optical fields with complex amplitudes $\alpha = \beta = 2$ and squeezing parameter $G = 0.5$. Fig. \ref{fig:fig4}(a) shows the time evolution of the uncertainty of the phase shift $\delta\phi$. The spectral density is $J(\omega)=\gamma\omega e^{-\frac{\omega}{\omega_c}}$ and the time-dependent function is $u_l=e^{-i\omega_l t}$. As shown in the Fig. \ref{fig:fig4}(a), $\delta\phi$ exhibits oscillatory behavior. When $\phi=2\Omega t=n\pi$, $n=0,1,2,\dots$, $\delta\phi$ reaches its minimum value. In this case, the minimum value of $\delta\phi$ remains stable and does not fluctuate over time. The corresponding time evolution of $\frac{1}{k}$ is shown in Fig. \ref{fig:fig4}(b). From this figure, we can see that $\frac{1}{k}<1$ is attainable.

More generally, we consider the condition that $u_l(t)=e^{(-\kappa_l+i\omega_l)t}$, here, the constant $\Delta$ which is generally renormalized into $\omega_0$. As shown in Figs. \ref{fig:fig4}(c) and \ref{fig:fig4}(d), the minimum value of $\delta\phi$ diverges rapidly over time, and resulting in a significant degradation of metrology sensitivity compared to the SNL. This implies that, in this case, the desired phase information is irretrievably lost. This loss occurs because, under the Born-Markovian approximation, the environment exhibits no memory effects, leading to the information leaked to the environment becoming irretrievable. Consequently, the advantages of our proposed quantum gyroscope scheme are completely undermined under the Born-Markovian approximation, which is precisely what we aim to avoid. Therefore, we will next explore the non-Markovian case.

In the context of the non-Markovian regimes, our primary concern is the form of the solution for $u_l(t)$, as it determines the behavior of $\delta\phi$ during long-time evolution. As mentioned earlier, to form a bound state, the condition $\gamma\omega_c\varGamma (s)>\omega_l$ must be satisfied. Therefore, for the case of $s=1$, $\gamma=0.05$ and $\Omega=0.1\omega_0$, we have the critical point: $\frac{\omega_c}{\omega_l}=20$. For $l=1$, we have $\frac{\omega_c}{\omega_0}=22$, and for $l=2$, we have $\frac{\omega_c}{\omega_0}=18$. Therefore, the solutions of $u_l(t)$ can be divided into three regions: without bound state when $\frac{\omega_c}{\omega_0}<18$, one bound state when $18<\frac{\omega_c}{\omega_0}<22$, and two bound states when $\frac{\omega_c}{\omega_0}>22$. The results of numerical simulation are shown in the Fig. \ref{fig:fig5}(a) to Fig. \ref{fig:fig5}(c). As illustrated in Fig. \ref{fig:fig5}(a) and Fig. \ref{fig:fig5}(b), in the absence of a bound state, the solution for $u_l(t)$ resembles that predicted by the Born-Markovian approximation, exhibiting a decay toward zero in the long-time limit. This behavior leads to the divergence of the minimum value of $\delta\phi$. Consequently, the subsequent analysis will concentrate on the scenario where the condition $\frac{\omega_c}{\omega_0}>22$ is satisfied.

\begin{figure}[htbp]
    \centering
    \begin{minipage}[t]{0.48\textwidth}
    \centering
        \includegraphics[width=\textwidth]{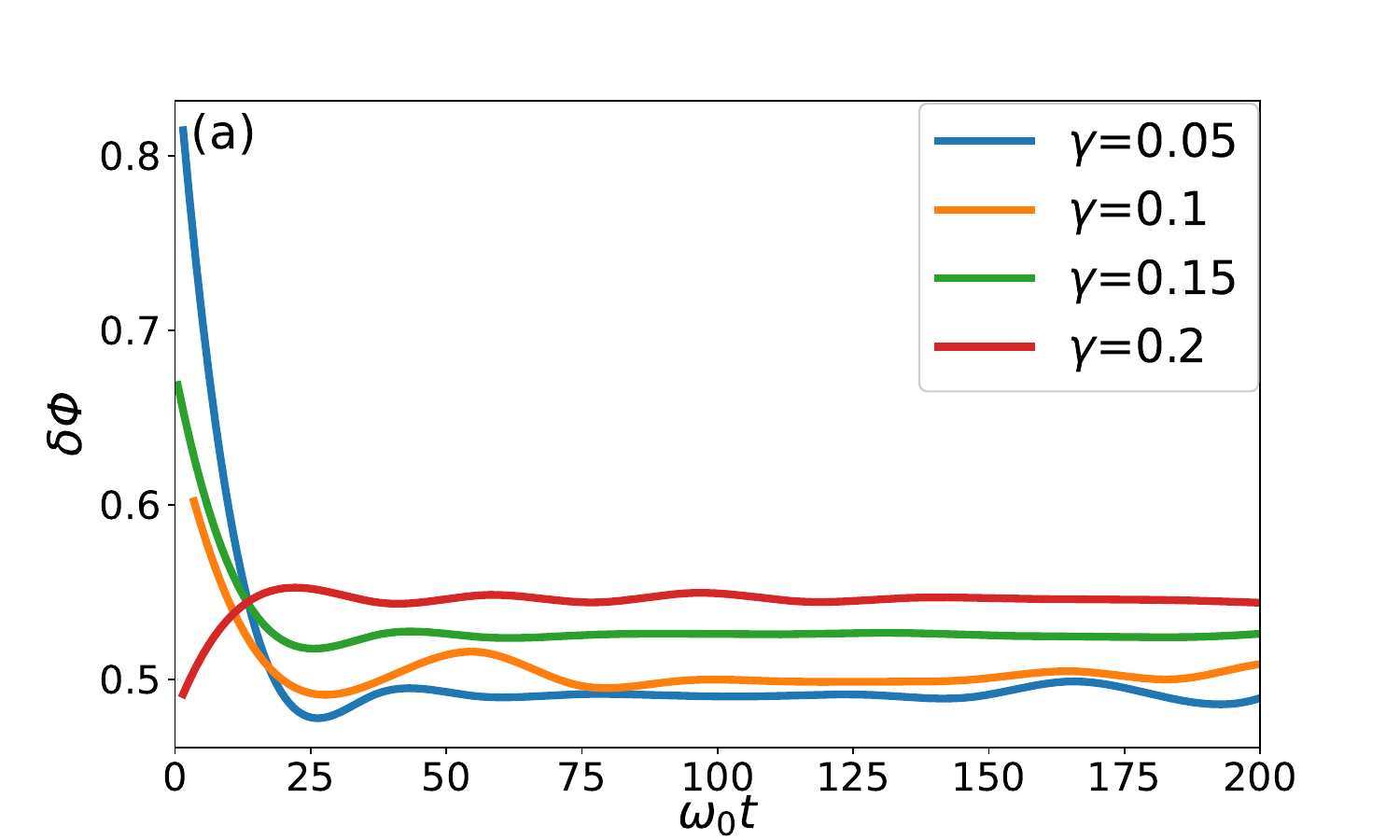}

        \label{fig:sub1}
    \end{minipage}\hfill
    \begin{minipage}[t]{0.48\textwidth}
    \centering
        \includegraphics[width=\textwidth]{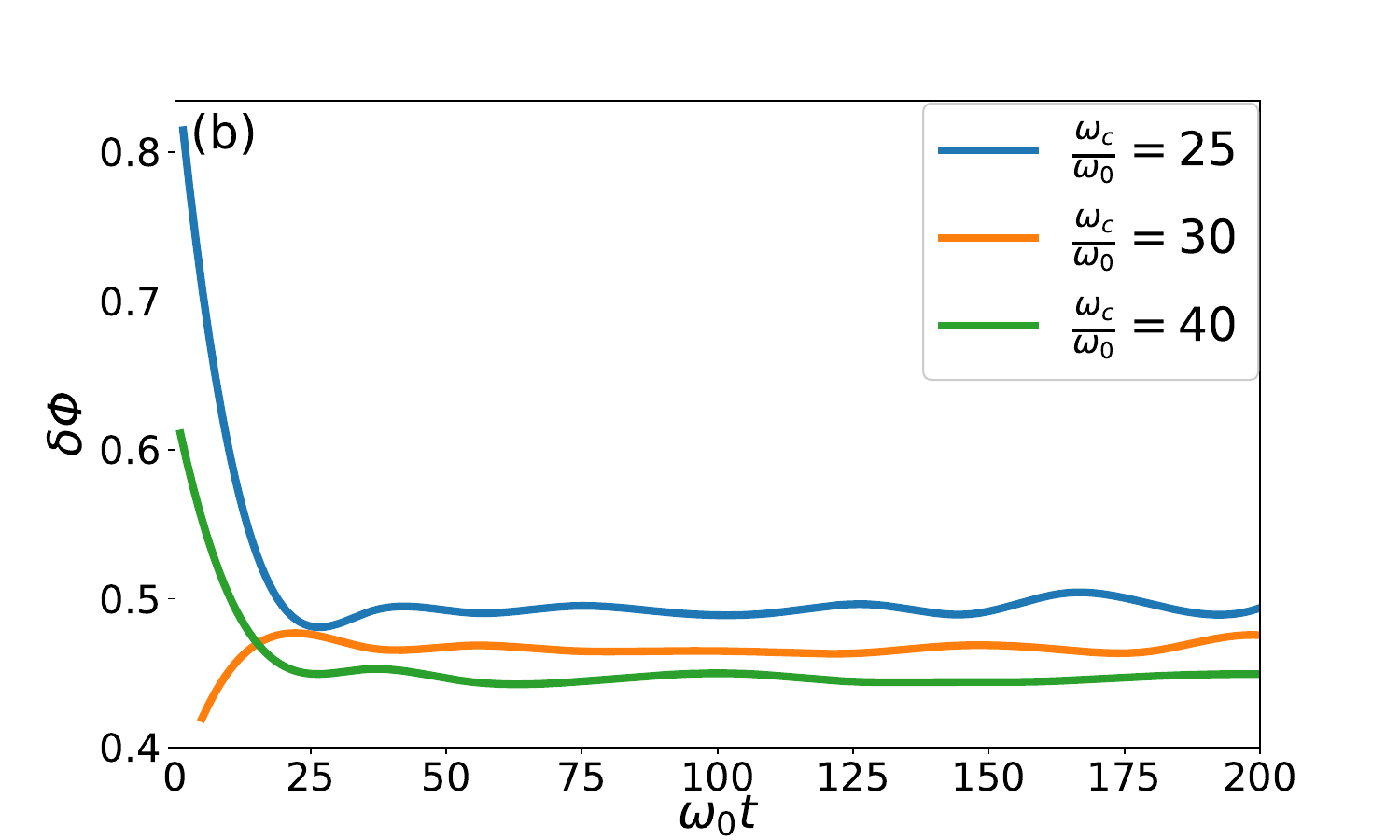}

        \label{fig:sub2}
    \end{minipage}

    \vspace{1em}

    \begin{minipage}[t]{0.48\textwidth}
    \centering
        \includegraphics[width=\textwidth]{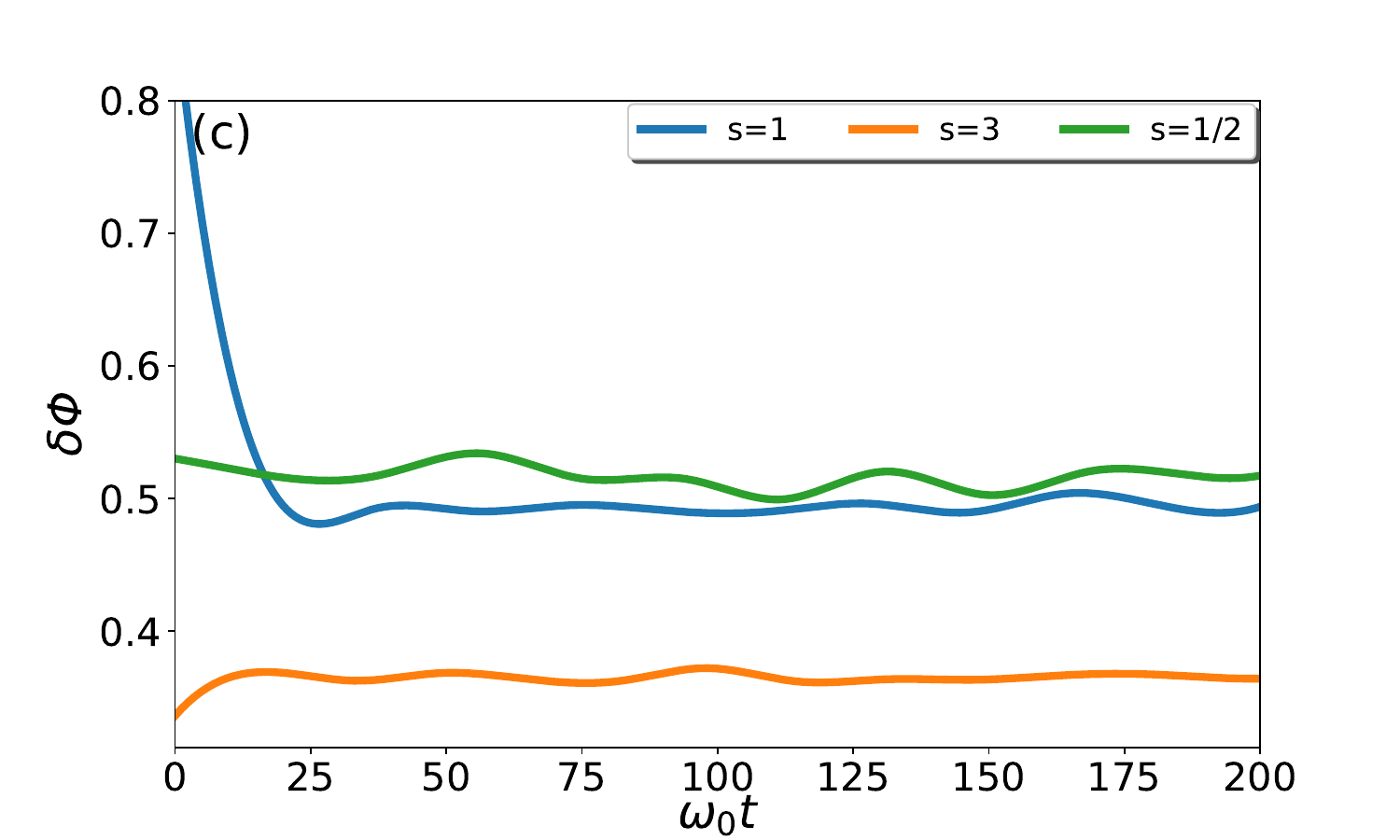}

        \label{fig:sub3}
\end{minipage}

    \caption{Local minima of $\delta\phi$ as a function of time for different values of (a) $\gamma$, (b) $\omega_c/\omega_0$, and (c) $s$. The parameters are: (a) $s=1$, $\Omega=0.1\omega_0$, $\omega_c/\omega_0=25$; (b) $s=1$, $\Omega=0.1\omega_0$, $\gamma=0.05$; and (c) $\gamma =0.05$, $\Omega=0.1\omega_0$, $\omega_c/\omega_0=25$.}
    \label{fig:fig6}
\end{figure}

Fig. \ref{fig:fig5}(c) illustrates the solution of $u_l(t)$ under this condition. It can be observed that as $t \to \infty$, $u_l(t)$ does not decay to zero but instead oscillates around a constant value. In this case, the minimum value of $\delta\phi$ does not diverge over time but oscillates within a very narrow range, as illustrated in Figs. \ref{fig:fig5}(d) and \ref{fig:fig5}(e). This suggests that by adjusting the energy spectrum of the total probe-environment system, the precision of the quantum gyroscope can be maintained in a stable state over time, rather than progressively deteriorating as time evolves.
When two bound states are formed, the Born-Markovian approximation breaks down, and non-Markovian effects allow the system to retain the information of interest. This effectively suppresses the decoherence caused by the environment.

It is essential to explore the the impact of the parameter $\gamma$ on the time evolution of the uncertainty in the phase shift $\delta\phi$. Fig. \ref{fig:fig6}(a) shows the evolution of the local minima of $\delta\phi$ for different values of $\gamma$. It is evident that, in the long-time limit, a larger $\gamma$ leads to a larger minimum value of $\delta\phi$. This result is quite intuitive, because a larger $\gamma$ indicates stronger coupling between the system and the environment. This intensified interaction facilitates greater dissipation, consequently increasing the minimum value of $\delta\phi$.
Fig. \ref{fig:fig6}(b) illustrates the effects of different $\frac{\omega_c}{\omega_0}$ on the system. To understand the influence of $\omega_c$ on the non-Markovian effects of the system, we first revisit the expression of the memory kernel $f(t-\tau)=\int_0^\infty J(\omega)e^{-i\omega (t-\tau)} \, d\omega$. When $\omega_c$ decreases, the dependence of the memory kernel on $t-\tau$ diminishes. In fact, as $\omega_c$ approaches zero, the memory kernel also approaches zero and becomes independent of $t-\tau$. Conversely, when $\omega_c$ increases, the dependence of the memory kernel on $t-\tau$ becomes stronger, and the memory effect of the environment becomes more pronounced. Therefore, the larger the value of $\omega_c$, the stronger the non-Markovian effect of the system. This is reflected in Fig. \ref{fig:fig6}(b) as an increase in the minimum value of $\delta\phi$; conversely, a larger $\omega_c$ would result in the opposite effect.

Finally, let us explore how different spectral density functions affect the system. We consider the cases of $s=1$, $s=\frac{1}{2}$, and $s=3$, which correspond to Ohmic, sub-Ohmic, and super-Ohmic spectra, respectively. As illustrated in Fig. \ref{fig:fig6}(c), the non-Markovian effect is most pronounced for the super-Ohmic spectrum. Similar to our analysis of $\omega_c$, the super-Ohmic spectrum encompasses a broader frequency range, which causes the memory kernel to decay more slowly, thereby exhibiting stronger non-Markovian effects.

\section{DISCUSSION AND CONCLUSIONS}
In order to provide a comprehensive overview of the experimental feasibility of the proposed gyroscope, it is first necessary to consider the theoretical underpinnings of the concept. Specifically, a spinning YIG microsphere affixed to a rotating platform has been experimentally realised in a recent report \cite{Maayani2018}. As demonstrated by Aspelmeyer et al. \cite{RevModPhys.86.1391}, a microsphere facilitates the support of two counter-circulating optical whispering gallery modes (WGMs), the excitation of which is significantly enhanced by the input light passing through a high-index prism. At the same time, when the input light is applied to the WGMs, the WGMs can be modulated by mechanical breathing \cite{RevModPhys.86.1391} and magnetization precession \cite{Chai:22}. Recent experimentation has demonstrated that the frequency of magnons in cavity magnonic systems generally ranges from a few hundred MHz up to approximately 50 GHz \cite{PhysRevLett.123.107702,PhysRevLett.113.156401}. Meanwhile, it has been demonstrated that spectral density is not only measurable \cite{10.1063/1.4900512,SALARISEHDARAN2019126006} but also tunable through the use of engineered reservoirs \cite{Myatt2000,doi:10.1126/science.1261033}, photonic crystal structures \cite{Liu2017}, and filtering pulse sequences \cite{Bylander2011}. The parameters employed in our numerical simulations are widely adopted in theoretical analyses and have been readily achieved experimentally ($\alpha = 2$ and $G = 0.5$) \cite{Yoon2024,YE2020126103,HOU20155102,PhysRevLett.98.030502}.  Consequently, it is theoretically to predict that, in the future, the quantum gyroscope may be achieve through specialized design.

In conclusion, we have demonstrated a cavity magnomechanics-based quantum gyroscope leveraging two-mode squeezed coherent states. We examined phase sensitivity under three distinct conditions and discovered that the highest sensitivity for the gyroscope is attained when the complex amplitudes of the two-mode coherent states are equal. Moreover, we found that in a dissipative environment, adjusting the relevant parameters of the environmental spectrum and introducing the non-Markovian effects can enable the gyroscope to maintain high precision even during long-term operation. Compared to previous quantum gyroscopes, the proposed quantum gyroscopes offer several significant advantages. Firstly, the size of the proposed models, which is cubic sub-millimeter in scale, fundamentally distinguishes it from conventional cubic-meter-scale quantum gyroscope. This distinction is crucial for their practical application in the field of navigation. Secondly, the proposed quantum gyroscopes have demonstrated a high degree of precision, which can be further enhanced by a factor of $e^{-G}$. Thirdly, the utilization of a two-mode coherent state as an input resource provides advantages in terms of ease of preparation, classical scalability, and experimental feasibility. Thus, our proposed gyroscope presents an effective solution for precise angle measurement in environments characterized by strong noise. We believe this scheme will greatly contribute to the advancement of high-precision rotation sensing.

This work was supported by the National Natural Science
Foundation of China under Grant 12074070.

\section*{APPENDIX:EXPECTATION VALUE OF INTENSITY DIFFERENCE OPERATOR}
\hypertarget{sec:intro}{}
For Eq. (\ref{eq37}), we will use Vernon's influence-functional theory to derive the evolution of the system's density matrix. The reduced density matrix of the system can be expressed as:
\begin{align}
\rho(\bm{\bar{\alpha}}_f,\bm{\alpha_f^{'}};t)=&\int d\mu(\bm{\alpha_i}) d\mu(\bm{\alpha_i^{'}})\mathcal{J}(\bm{\bar{\alpha}}_f,\bm{\alpha_f^{'}};t|\bm{\bar{\alpha}}_i,\bm{\alpha_i^{'}};0)\nonumber\\
&\times \rho(\bm{\bar{\alpha}}_i,\bm{\alpha_i^{'}};0),
\end{align}
\begin{align}
\mathcal{J}(\bm{\bar{\alpha}}_f,\bm{\alpha_f^{'}};t|\bm{\bar{\alpha}}_i,\bm{\alpha_i^{'}};0) = & \exp \left\{ \sum_{l=1}^{2} \left\{ u_l(t)\bm{\bar{\alpha}_{lf}}\bm{\alpha}_{li} \right. \right. \nonumber \\
& \left. + \bar{u}_l(t)\bm{\bar{\alpha}_{li}^{'}}\bm{\alpha}_{lf}^{'} \right. \nonumber \\
& \left. \left. + [1-|u_l(t)|^2]\bm{\bar{\alpha}_{li}^{'}}\bm{\alpha}_{li} \right\} \right\},\label{eq44}
\end{align}
where $\rho(\bm{\bar{\alpha}}_f,\bm{\alpha_f^{'}};t)=\langle\bm{\bar{\alpha}_f}|\rho(t)|\bm{\alpha_f^{'}}\rangle$ is the reduced density matrix expressed in coherent-state representation and $\mathcal{J}(\bm{\bar{\alpha}}_f,\bm{\alpha_f^{'}};t|\bm{\bar{\alpha}}_i,\bm{\alpha_i^{'}};0)$ is the propagating function. Here, we have $|\bm{\alpha}\rangle=\prod_{l=1}^{2}|\bm{\alpha_l}\rangle$, $\hat{a}_l|\bm{\alpha_l}\rangle=\bm{\alpha_l}|\bm{\alpha_l}\rangle$, and $\langle\bm{\bar{\alpha}}|\bm{\alpha^{'}}\rangle=\exp(\bm{\bar{\alpha}}\bm{\alpha^{'}})$. The integration measure $d\mu(\bm{\alpha})=\prod_{l}e^{-\bm{\bar{\alpha}_l}\bm{\alpha_l}}\frac{d\bm{\bar{\alpha}_l}d\bm{\alpha_l}}{2\pi i}$. $\bm{\bar{\alpha}}$ is the complex conjugate of $\bm{\alpha}$.
The propagating function $\mathcal{J}(\bm{\bar{\alpha}}_f,\bm{\alpha_f^{'}};t|\bm{\bar{\alpha}}_i,\bm{\alpha_i^{'}};0)$ is given by Eq. (\ref{eq44}) \cite{AN20091737}, and $u_l(t)$ satisfies Eq. (\ref{eq38}).

The initial input state is a two-mode squeezed coherent state $|\Phi_{in}\rangle=S(G)|\alpha\rangle |\beta\rangle$, where G is the squeezing parameter, and for simplicity in calculations, here we consider the case where $\alpha$ = $\beta$. In the coherent-state representation, this initial state is given by

\begin{align}
\rho(\bm{\bar{\alpha}}_f,\bm{\alpha_f^{'}};0)=&\frac{1}{\cosh^2G}\exp[\frac{-i\tanh G}{2}(\bm{\bar{\alpha}_i}^2-{\bm{\alpha_i}^{'}}^2-2\bm{\bar{\alpha}_i}\tilde{\bm{\alpha}}\nonumber\\
+&2\bm{\alpha}_i^{'}\tilde{\bm{\alpha}})+\frac{1}{2}(\tilde{\bm{\alpha}}\bm{\alpha_i}-\tilde{\bm{\alpha}}^{*}\bm{\bar{\alpha}_i}-\tilde{\bm{\alpha}}\bm{\bar{\alpha_i}}^{'}+\tilde{\bm{\alpha}}^{*}\bm{\alpha_i}^{'})],
\end{align}
where $\bm{\tilde{\alpha}}=\bm{\alpha}[\cosh(G)-\sinh(G)]$ and $\bm{\tilde{\alpha}}^{*}$ is the complex conjugate of $\bm{\tilde{\alpha}}$. The time-dependent reduced density matrix is obtained by Eq. (44), and the final density matrix can be expressed as:
\begin{align}
\rho_{out}=&\int d\mu(\bm{\alpha_f})d\mu(\bm{\alpha_f}^{'})\rho(\bm{\bar{\alpha}}_f,\bm{\alpha_f^{'}};t)V|\bm{\alpha_{1f}},\bm{\alpha_{2f}}\rangle \cdot \nonumber\\ &\langle\bm{\bar{\alpha_{1f}}^{'}},\bm{\bar{\alpha_{2f}}^{'}}|V^{\dagger},
\end{align}
where $V=\exp[i \frac{\pi}{4}(a^{\dagger}m+am^{\dagger})]$. Then the expectation value $\langle n_{d} \rangle = Tr[n_{d}\rho_{out}]$ of the intensity difference operator $n_{d}=b_1^{\dagger} b_1 - b_2^{\dagger} b_2$ can be calculated as
\begin{widetext}
\begin{align}
\langle n_{d} \rangle=\frac{ix(B_1-B_2)exp[\frac{m_1q_1^2+\bar{m}_1n_1^2+n_1q_1}{(p_1-1)^2-4|m_1|^2}+\frac{m_2q_2^2+\bar{m}_2n_2^2+n_2q_2}{(p_2-2)^2-4|m_2|^2}+e_1+e_2]}{[(p_2-1)^2-4|m_2|^2]^{\frac{1}{2}}[(p_2-1)^2-4|m_2|^2]^{\frac{1}{2}}},\nonumber
\end{align}
\begin{equation}
B_1=\frac{[(n_1 + q_1)(2\bar{m}_1 + 1 - p_1) + (n_1 - q_1)(2\bar{m}_1 - 1 + p_1)][(n_2 + q_2)(2m_2 + 1 - p_2) - (n_2 - q_2)(2m_2 - 1 + p_2)]}{4[(p_1-1)^2-4|m_1|^2][(p_2-1)^2-4|m_2|^2]},\nonumber
\end{equation}
\begin{equation}
B_2=\frac{[(n_1 + q_1)(2m_1 + 1 - p_1) + (q_1 - n_1)(2m_1 - 1 + p_1)][(n_2 + q_2)(2\bar{m}_2 + 1 - p_2) - (q_2 - n_2)(2\bar{m}_2 - 1 + p_2)]}{4[(p_1-1)^2-4|m_1|^2][(p_2-1)^2-4|m_2|^2]},\nonumber
\end{equation}
\begin{equation}
q_l=\frac{\frac{\tanh^2(G)}{2}(1-|u_l|^2)\tilde{\alpha}\bar{u}_l+i\tanh(G)\tilde{c}_l\bar{u}_l-c\bar{u_l}}{A_l},\nonumber
\end{equation}
\begin{equation}
n_l=\frac{i\tanh(G)u_l(1-|u_l|^2)c-\frac{i\tanh^3(G)(1-|u_l|^2)^2\tilde{\alpha}u_l}{2}+\tanh^2(G)(1-|u_l|^2)u_l\tilde{c}_l}{A_l}-\frac{i\tanh(G)}{2}u_l\tilde{\alpha}+cu_l,\nonumber
\end{equation}
\begin{equation}
\begin{split}
e_l&=\frac{\frac{\tanh^2(G)}{2}(1-|u_l|^2)\tilde{\alpha}\tilde{c}_l+\frac{i\tanh(G)}{2}\tilde{c}^2_l+\frac{i\tanh(G)c(1-|u_l|^2)\tilde{\alpha}}{2}-\frac{i\tanh(G)}{2}(1-|u_l|^2)^2c^2-\frac{i\tanh^3(G)}{8}
(1-|u_l|^2)^2\tilde{\alpha}^2-c\tilde{c}_l}{A_l}\\
& \quad+\frac{c\tilde{\alpha}}{2}-\frac{i\tanh(G)\tilde{\alpha}^2}{8},
\end{split}
\end{equation}
\end{widetext}
where $x=(\sqrt{A_1A_2}\cosh^2 G)^{-1}$, $m_l=\frac{-iu_l(t)^2\tanh G}{2A_l}$, $p_l=|u_l(t)|^2(1-A_l)^{-1}$, $A_l=1-[|u_l(t)|^2-1]^2\tanh^2 G$, $c=(\frac{\tilde{\alpha}}{2}-i\tilde{\alpha}\tanh G)^2$, and $\tilde{c}_l=c(1-|u_l|^2)-\frac{\tilde{\alpha}}{2}$. The sensing sensitivity of $\phi$ is calculated by $\delta\phi=\frac{\sqrt{\langle n_{d}^2 \rangle - {\langle n_{d} \rangle}^2}}{|\partial_\phi \langle n_d \rangle|}$.
\bibliography{citation}
\end{document}